\documentclass[twocolumn]{aastex62}

\graphicspath{{./}{figures/}}

\received{\today}
\revised{\today}
\accepted{\today}

\submitjournal{ApJ}
\shorttitle{SNTF total-intensity}
\shortauthors{Par\'e et al.}

\begin{document}

\title{A VLA Study of Newly-Discovered Southern Latitude Non-Thermal Filaments in the Galactic Center: Radio Continuum Total-intensity and Spectral Index Properties}

\correspondingauthor{Dylan Par\'e}
\email{dylanpare@gmail.com}

\author[0000-0002-5811-0136]{Dylan M. Par\'e}
\affil{Villanova University \\
800 East Lancaster Avenue \\
Mendel Hall \\
Villanova, PA 19085}

\author{Cornelia C. Lang}
\affil{University of Iowa \\
30 North Dubuque Street, Room 203 \\
Iowa City, IA 52242}

\author[0000-0002-6753-2066]{Mark R. Morris}
\affil{University of California, Los Angeles \\
430 Portola Plaza, Box 951547 \\
Los Angeles, CA 90095-1547}

\begin{abstract}

The non-thermal filament (NTF) radio structures clustered within a few hundred parsecs of the Galactic Center (GC) are apparently unique to this region of the Galaxy. Recent radio images of the GC using MeerKAT at 1 GHz have revealed a multitude of faint, previously unknown NTF bundles (NTFBs), some of which are comprised of as many as 10 or more individual filaments. In this work we present Very Large Array (VLA) observations at C- and X-bands (4 - 12 GHz) at arcsecond-scale resolutions of three of these newly-discovered NTFBs, all located at southern Galactic latitudes. These observations allow us to compare their total-intensity properties with those of the larger NTF population. We find that these targets generally possess properties similar to what is observed in the larger NTF population. However, the larger NTF population generally has steeper spectral index values than what we observe for our chosen targets. The results presented here based on the total-intensity properties of these structures indicate that the NTFs are likely all formed from Cosmic Rays (CRs). These CRs are either generated by a nearby compact source and then diffuse along the NTF lengths or are generated by extended, magnetized structures whose magnetic field undergoes reconnection with the NTF magnetic field.
\end{abstract}

\section{INTRODUCTION} \label{sec:intro}

The Galactic Center (GC) is a nearby galactic nuclear region (8.0 kpc away, \citealt{Abuter2019,Do2019}) that exhibits extreme properties compared to what is observed in the Galactic Disk, such as elevated magnetic field strengths (100s of $\rm\mu$G to mG, e.g. \citealt{YM1987,Ferriere2011,Pillai2015,Mangilli2019}). By studying the GC, we can infer properties of more distant galactic nuclear regions if we assume the GC is a representative nuclear region. A population of structures exists in the GC that appear as glowing threads at radio wavelengths. These structures are known as the non-thermal filaments (NTFs) and have to date only been observed in the GC \citep{Gray1995,Morris1996b,Yusef-Zadeh2004}.

The NTFs are highly polarized synchrotron sources that are 10s to 100s of times longer than they are wide \citep{Morris1996b}. The synchrotron emission from these sources reveals the presence of relativistic electrons \citep{Morris1996b}. However, the source and mechanism producing these relativistic electrons remains unclear (see \citealt{Morris1996c,Shore1999,Bicknell2001,Yusef-Zadeh2003,Yusef-Zadeh2019,Bykov2017,Sofue2020,Thomas2020} and references therein). 

The first NTF detected within the GC is also the most prominent and is known as the Radio Arc \citep{YMC1984,Yusef-Zadeh1986a,YM1987}. The Radio Arc consists of  $\rm\geq$10 individual filaments in close proximity to one another that are all oriented perpendicular to the Galactic Plane. Each filament is individually quite narrow, with widths of $\rm\sim$0.5$\rm\arcsec$ (0.02 pc, \citealt{Pare2019}). 

For multiple decades, the Radio Arc was the only NTF system known to consist of so many individual filaments. By comparison, other NTFs were generally revealed to consist of only a few filaments (e.g. \citealt{Gray1995,Lang1999a,Lang1999b}). However, recent observations using MeerKAT have uncovered numerous fainter NTF bundles (NTFBs) that have a similar morphology to the Radio Arc in that they consist of several filaments \citep{Thomas2020,Heywood2022,Yusef-Zadeh2022}. Though the Radio Arc exists in a complex region of the GC, the faint, newly-discovered NTFBs are located in seemingly calmer regions of the GC that are further removed from the central black hole, Sgr A$\rm^*$. The discovery of these fainter NTFBs begs the question of whether all NTFs in the GC have a shared formation and evolutionary history.

Further motivating the question of whether the NTFs consist of multiple distinct populations are the recent results obtained in \citet{Pare2019}, which reveal that the Radio Arc possesses an alternating magnetic field pattern which varies from being parallel to the NTFB orientation to being rotated by a 60 degree angle with respect to the NTFB orientation. All other NTFs are shown to possess generally parallel magnetic fields which trace the NTF orientation \citep{Gray1995,Lang1999a,Lang1999b}. Follow-up studies of the Radio Arc have indicated that the unusual magnetic field pattern could be a result of a magnetized medium local to the Radio Arc, rather than being an intrinsic property of the Radio Arc itself \citep{Pare2021}.

In this paper, we analyze observations taken with the Karl G. Jansky Very Large Array (VLA) managed by the NRAO\footnote{The National Radio Astronomy Observatory is a facility of the National Science Foundation operated under cooperative agreement by Associated Universities, Inc.} of three faint, newly-discovered NTFBs in the GC that are detected at various locations within the region. By comparing the properties of these new NTFBs which are morphologically similar to the Radio Arc with the general NTF population we aim to demonstrate the degree to which these NTF populations are related to one another. 

We focus on the comparison of total-intensity features between our targets and previously-studied NTFs by analyzing the environments around our targets, their subfilamentation, and their spectral indices and compare these properties with those seen for the other NTFs in the GC.

In Section \ref{sec:dat_red} we detail the data calibration and imaging process implemented to produce the data used in this paper. In Section \ref{sec:res} we summarize the results obtained from the observations of our targets. Section \ref{sec:disc} presents the implications of our results. We conclude the paper in Section \ref{sec:conc} by summarizing the key points revealed by our analysis.

\section{OBSERVATIONS AND DATA REDUCTION} \label{sec:dat_red}

\subsection{Observations} \label{sec:obs}
\begin{deluxetable*}{|c|c|c|c|c|c|c|}[ht!]
\tablecaption{Summary of Southern Latitude NTFB VLA Observations}
\tablecolumns{7}
\tablenum{1}
\tablewidth{0pt}
\tablehead{
\colhead{Object} & \colhead{Freq. Band} & \colhead{Freq. Range (GHz)} & \colhead{T$\rm_{int}$/field (min)} & \colhead{N$\rm_{fields}$} & \colhead{Array Configs.} & \colhead{Total Time (hrs)}
}
\startdata
SNTF1 & C & 4.0 - 8.0 & 50 & 2 & B, C & 1.67 \\
      & X & 8.0 - 12.0 & 45 & 3 & C & 1.5 \\
SNTF2 & C & 4.0 - 8.0 & 50 & 3 & B, C & 2.5 \\
      & X & 8.0 - 12.0 & 45 & 6 & C & 3.0 \\
SNTF3 & C & 4.0 - 8.0 & 50 & 3 & B, C & 2.5 \\
      & X & 8.0 - 12.0 & 45 & 5 & C & 2.50 \\
\enddata
\tablecomments{Object indicates the name of the NTFB observed, Freq. band denotes the name of the frequency band used for the observation, Freq. Range shows the frequency range of the frequency band, T$\rm_{int}$/field shows the amount of time observed for each pointing with the same amount of time being used to observe the targets in B- and C-array, N$\rm_{fields}$ shows the number of fields observed and mosaicked together, Array Configs. details the VLA array configurations used for each frequency band, and Total Time shows the total time observed for each frequency band.}
\end{deluxetable*}

Three NTFBs were observed at both C- and X-bands using the B- and C-configurations of the VLA during the Spring of 2020. We refer to these targets throughout the paper as SNTF1, SNTF2, and SNTF3 (``S'' for Southern Galactic Latitudes). We used multiple pointings for each target sufficient to fully image each NTFB at the two frequency bands observed. The details of the observations are shown in Table 1. Some bright compact sources located near the edge of the primary beam had the potential to produce significant imaging artifacts; however, using widefield imaging, we determined that the bordering point sources had negligible effects on our targets.

\subsection{Calibration} \label{sec:cal}

Three calibrators were observed for each frequency band. 3C-286 was used as the flux and bandpass calibrator, J1407+284 was used as the polarization calibrator, and 1751-253 was used as the phase calibrator. 3C-286 and J1407+284 were observed once at the beginning of each observation for each of the NTFBs, whereas 1751-253 was observed intermittently throughout the observations, roughly once every 8 minutes.

To calibrate our NTFB observations we used the Common Astronomy Software Applications (CASA) VLA Calibration Pipeline\footnote{https://science.nrao.edu/facilities/vla/data-processing/pipeline}. This pipeline flags the data for intensity spikes and other Radio Frequency Interference (RFI) and implements common total-intensity calibration steps like flux and bandpass calibration. However, the pipeline does not currently support polarization calibration. Since only the total-intensity data is analyzed in this paper, the calibration performed by the pipeline is sufficient for the data presented here. Our polarization calibration is detailed in paper II, where we present the polarized intensity results (Par\'e et al. 2023, \textit{in prep}). Running the standard calibration pipeline produced a high-quality calibration of the total-intensity data and no follow-up calibration utilizing more sophisticated calibration methods were needed for this work.

\subsection{Imaging} \label{sec:imag}

2D total-intensity distributions were used to analyze the NTFB environments, morphologies, and spectral indices. We used a Multi-Frequency Synthesis (MFS) cleaning routine to produce the 2D total-intensity distributions of our targets. This deconvolution method produced one intensity distribution for each of our frequency bands, C-band (6 GHz) and X-band (10 GHz), following the procedure detailed in \citep{Clark1980}. We used a Briggs weighting of 0.5 and 100,000 cleaning iterations. A primary beam correction was also implemented to account for the reduced sensitivity of the interferometer near the edge of the field-of-view.  This procedure was implemented using the CASA task \textit{tclean}, which was run separately for each frequency band and target observed. 
We mosaicked the individual pointings into one combined image of each source at each frequency band, and the resulting mosaics at 6 and 10 GHz are presented and described in Section \ref{sec:res}.

The widths of the NTFBs are determined using these images in two different ways. The observed width was determined through inspection of the observed filament sizes from direct inspection of  the total intensity images. In addition, a quadrature decomposition was employed to correct the observed filament widths with the beam size perpendicular to the orientation of the filament: $\rm{}IW = \sqrt{W^2 - BW^2}$ where IW is the intrinsic width of the filament (arcsec), W is the observed width of the filament (arcsec), and BW is the width of the beam size of the synthesized beam in the direction perpendicular to the NTFs (arcsec). Both the observed and intrinsic widths for the NTFBs are discussed in Section \ref{sec:res}.

\subsection{Spectral Index Derivation} \label{sec:a_deriv}
Deriving the spectral indices of our sources allows us to compare the spectral properties of our NTFBs with those of previously-observed NTFs. For this work we define the spectral index as $S_{\nu} \propto \nu^{\alpha}$ where $\rm\nu$ is the frequency of observation, $\rm{}S_{\nu}$ is the intensity of the source at the observed frequency, and $\rm\alpha$ is the spectral index.

Before the spectral index can be determined we ensured that both the restoring beam size and pixel cell size are the same across our 6 and 10 GHz data sets. We also matched the \textit{uv}-spacing of the 6 and 10 GHz observations for a given NTFB. It is also important to ensure that our total intensity distributions in the different frequency bands are on the same spatial grid. These modifications ensure that the lines of sight used to determine the spectral index cover the same spatial extent and cover the same spatial location of the target. These modifications were implemented in CASA by enforcing a 5.0$\rm\arcsec$ circular restoring beam size and a 1\arcsec\ cell size at both 6 and 10 GHz during the cleaning process. The regridding procedure was implemented using the CASA task \textit{imregrid()} which aligns the spatial coordinate systems of the different distributions. 

It is also important to correct for the background emission when analyzing the spectral index of our sources; however, because of the rapidly varying background present in our observations we were unable to fully correct for the background contamination. To partially correct for the background, we subtracted an average background level from the distributions used to calculate the spectral index. This average background was calculated by extracting the average intensity obtained from background regions of our distributions.

Once we produce smoothed and regridded images we derive the spectral index of our targets by comparing the intensity detected at 6 GHz with the intensity observed at 10 GHz by fitting a straight line between these two measurements in log-log space. Because of our inability to fully correct for the spatial varying background contamination, we did not produce maps of the spectral index. Instead, we averaged the spectral index values derived for each line of sight for each 20$\rm\arcsec$ segment along the length of each NTFB. The plots of these average spectral index values as a function of NTFB length are shown in Section \ref{sec:res}.

\section{RESULTS} \label{sec:res}
\begin{figure*}
    \centering
    \plotone{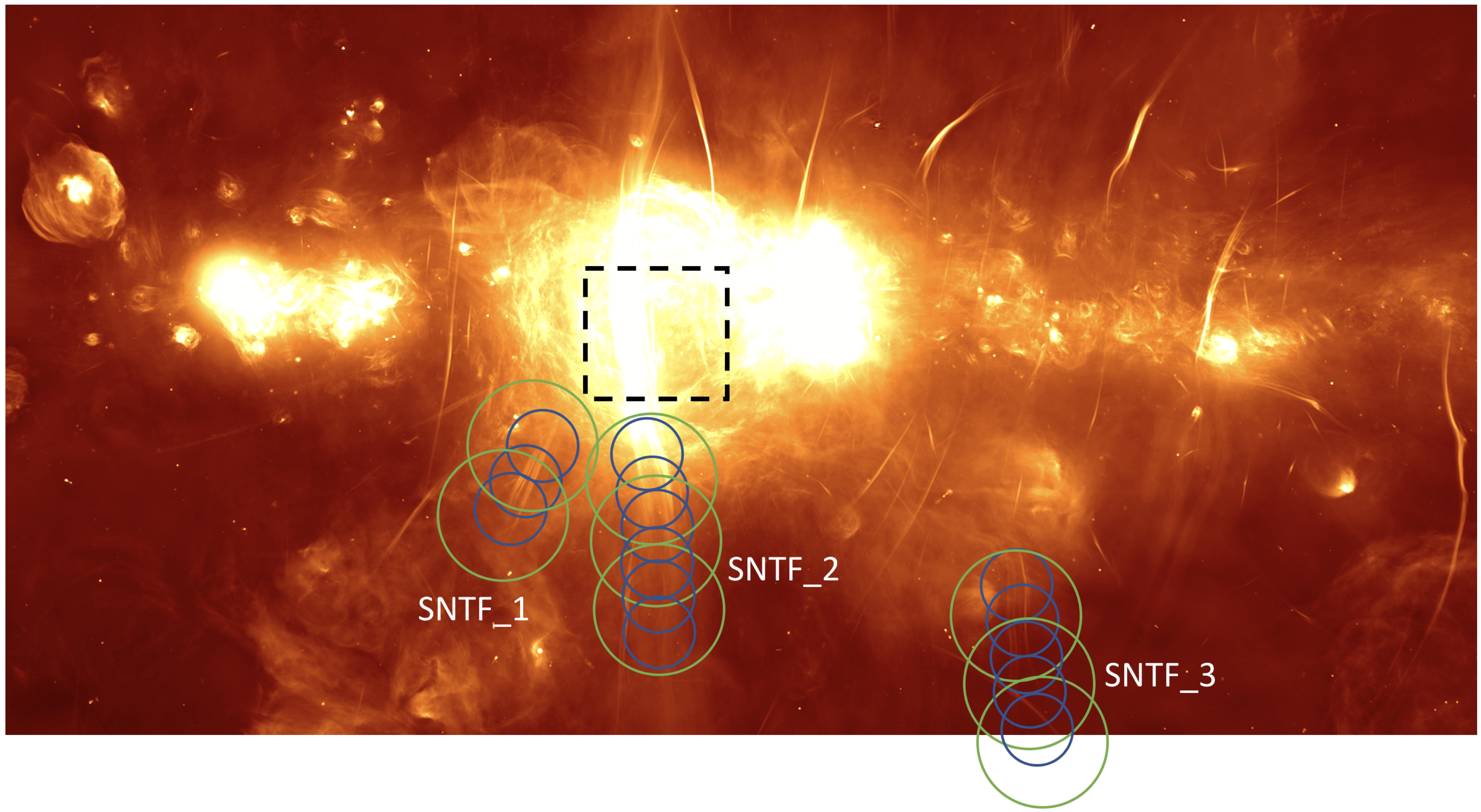}
    \caption{1 GHz MeerKAT image of the GC \citep{Heywood2022,Yusef-Zadeh2022} with the VLA fields of view for our NTFB observations overlayed as circles. X-band fields of view are in blue and C-band fields of view are in green. The black rectangle indicates the field of view of previous Radio Arc observations (e.g. \citealt{Pare2019,Pare2021}).}
    \label{fig:chart}
\end{figure*}
We observed three NTFBs for this work. The location of each of our targets is shown in a finding chart in Figure \ref{fig:chart}. The dashed black rectangle shown in Figure \ref{fig:chart} indicates the field of view of previous Radio Arc observations presented in \citet{Pare2019,Pare2021}. The NTFBs observed in this work are labeled SNTF1, SNTF2, and SNTF3 in this image, and we use this labeling throughout the rest of this paper to refer to our targets.

\subsection{SNTF1 total-intensity}
The first of our targets is located to the East of the Radio Arc. SNTF1 (G0.30-0.26) has an angular separation from the Radio Arc of $\rm\sim$5$\rm\arcmin$ (12 pc) and is located at southern latitudes within the GC, beginning roughly 8$\rm\arcmin$ ($\rm\sim$19 pc) below the Galactic Plane.

\begin{figure*}
    \centering
    \plotone{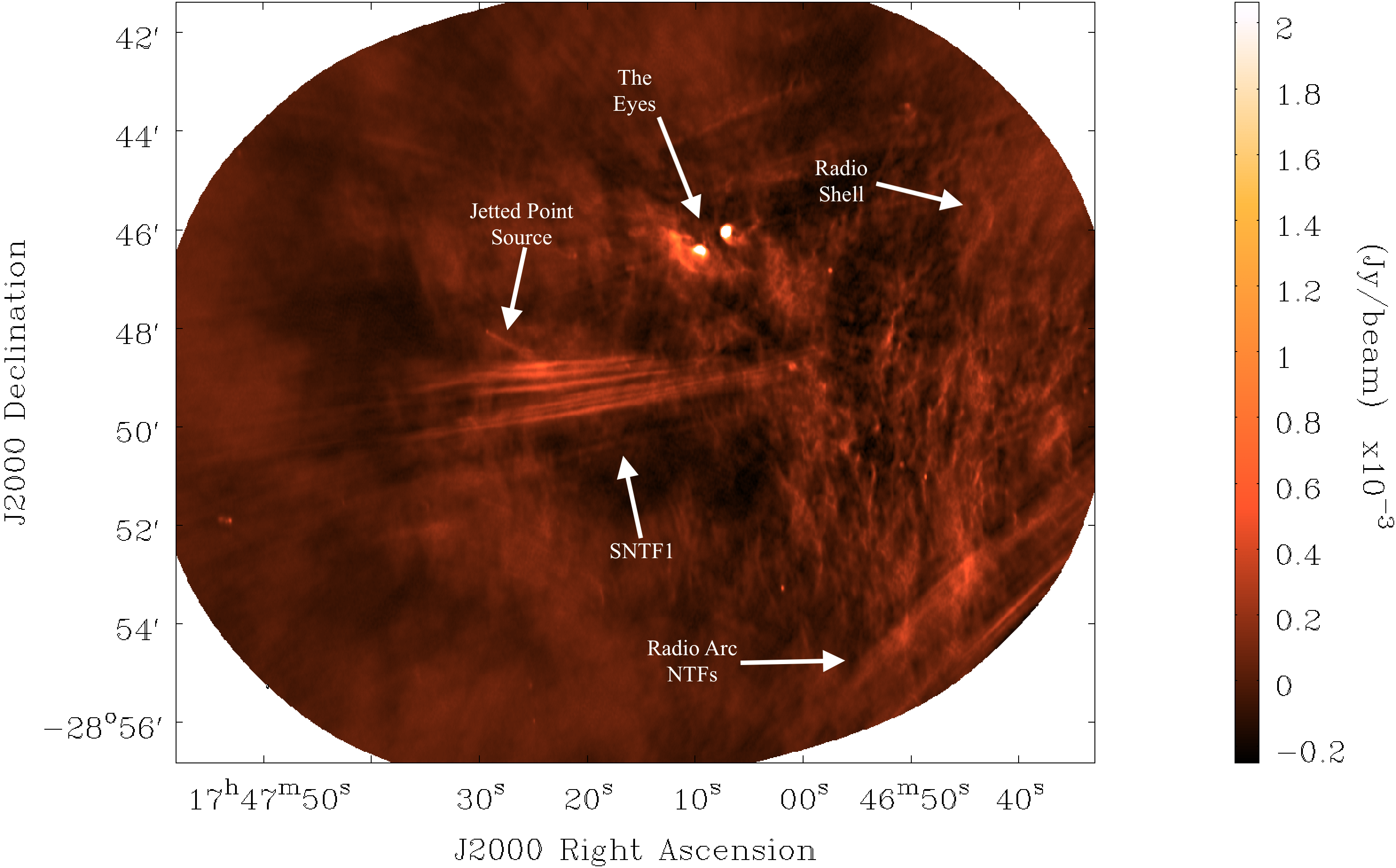}
    \caption{6 GHz total-intensity distribution of SNTF1 (G0.30-0.26). This image has a beam size of 4.1$\rm\arcsec$ x 2.1$\rm\arcsec$ and an rms noise level of $\rm\sim{}100\mu$Jy beam$\rm^{-1}$. Key features are labeled in the image.}
    \label{fig:SNTF1_C}
\end{figure*}
\begin{figure*}
    \centering
    \plotone{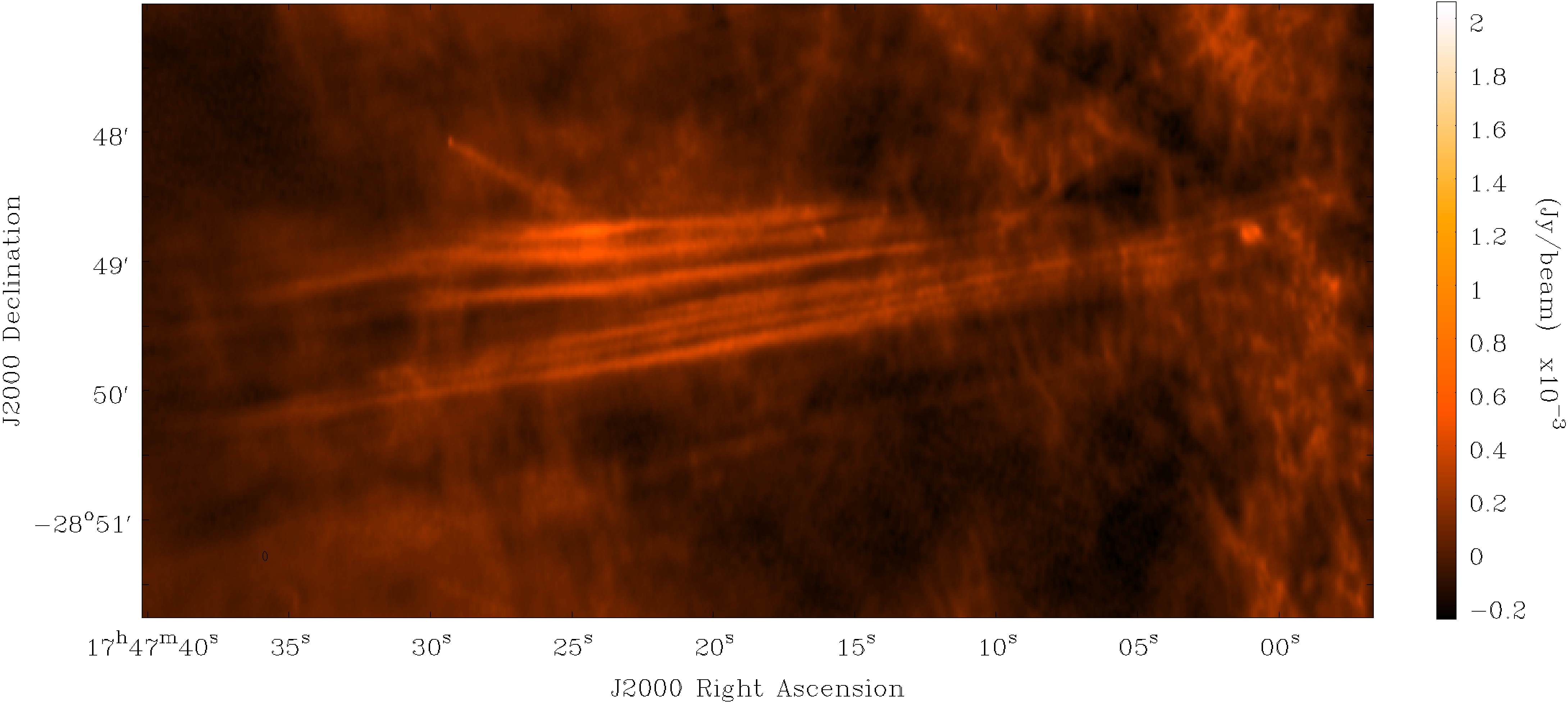}
    \caption{Zoomed--in view of SNTF1 (G0.30-0.26) at 6 GHz. Beam size and rms noise level are the same as in Figure \ref{fig:SNTF1_C}.}
    \label{fig:SNTF1_Zoom_C}
\end{figure*}
\begin{figure*}
    \centering
    \plotone{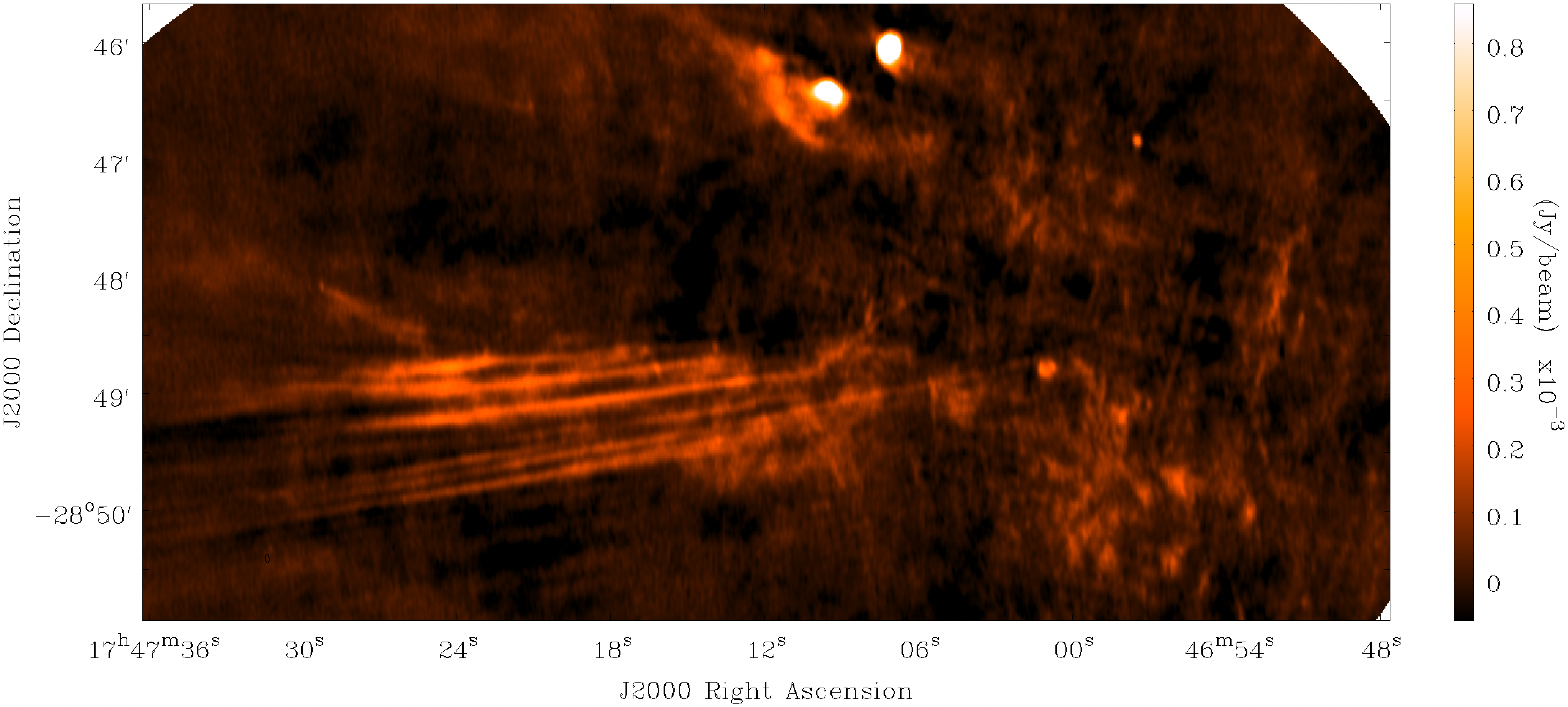}
    \caption{10 GHz total-intensity distribution of SNTF1 (G0.30-0.26). This image has a beam size of 4.7\arcsec\ x 1.8\arcsec\ and an rms noise level of $\rm\sim{}60\mu$Jy beam$\rm^{-1}$.}
    \label{fig:SNTF1_X}
\end{figure*}
Figure \ref{fig:SNTF1_C} shows the total-intensity distribution obtained for SNTF1 at a central frequency of 6.0 GHz. There are seven prominent individual filaments within SNTF1, with multiple fainter filaments also detected. This structure is also observed in \citet{Yusef-Zadeh2004}, \citet{Heywood2022}, and \citet{Yusef-Zadeh2022} (labeled as feature S3 in \citealt{Heywood2022,Yusef-Zadeh2022}). Though the morphology of SNTF1 is the same as seen at 1 GHz, we do not detect the oblique filament that crosses SNTF1 at 17:47:38, -28:50:15 as seen in \citet{Heywood2022}. The fact that this oblique filament does not appear in our higher frequency observations could indicate that it has a steeper spectral index than the other NTF structures observed in this work. However, since the oblique filament is also fainter than SNTF1 is observed at 1 GHz our observations may not have been sensitive enough to detect it.

A portion of the Radio Arc NTFB is visible in the southwestern portion of Figure \ref{fig:SNTF1_C}, as marked in the figure. There is also what appears to be an angled filament located at 17:47:28, -28:48:30 (G0.33-0.27) that intersects SNTF1. This angled filament is oriented at a 60 degree angle from SNTF1. There are also multiple additional point sources that can be identified in Figure \ref{fig:SNTF1_C}. Through the zoomed-in view of SNTF1 shown in Figure \ref{fig:SNTF1_Zoom_C} it is evident that the angled filament is actually a point source with an apparent tail or jet of emission. Alternatively, it could be a wake from the point source created as it moves away from SNTF1. This jet or wake approaches or recedes from the SNTF1 filaments and possibly extends into the structure.

Figure \ref{fig:SNTF1_X} shows the total-intensity distribution for SNTF1 at a central frequency of 10.0 GHz. Because of the smaller field of view at this higher frequency, the portion of the Radio Arc and most of the point sources observed at 6.0 GHz are not observed. At this higher frequency the morphology of the filaments within SNTF1 appear the same as what is observed at 6 GHz. The filaments comprising SNTF1 are all generally parallel to one another. The filaments have an average length of 6.6$\rm\arcmin$ (15 pc), and an average width of 3.0$\rm\arcsec$ (0.12 pc) as measured from the 10 GHz data shown in Figure \ref{fig:SNTF1_X}. We also perform a quadrature decomposition as discussed in Section \ref{sec:cal} and using a beam width of 2.0\arcsec\ we find an intrinsic width of 2.2\arcsec\ (0.09 pc). The length and width parameters for the filaments in SNTF1 are also displayed in Table 2, which displays the observed and derived properties of our NTFB targets.

While the brightnesses of the head-tail filament and of the observed portion of the Radio Arc are generally constant along their lengths, there is an apparent brightness enhancement in the SNTF1 filaments coinciding with the location where the angled filament crosses the SNTF1 filaments. This brightness enhancement is most easily seen at 6 GHz in Figure \ref{fig:SNTF1_Zoom_C}, but is also observed at 10 GHz as seen in Figure \ref{fig:SNTF1_X}. There is also a bifurcation in the SNTF1 system that occurs at 17:47:23, -28:49:21 from where the individual filaments progressively concentrate toward the East into two distinct clumps. This bifurcation coincides with the location in SNTF1 at which the angled filament crosses the SNTF1 filaments.

SNTF1 is observed near two bright compact sources located at 17:47:09.43, -28.46.25.3 and 17:47:07.17, -28:46:02.2 (G0.31-0.20, G0.31-0.19), labeled as ``The Eyes'' in Figure \ref{fig:SNTF1_C} and also seen in Figure \ref{fig:SNTF1_X}. One of these compact sources  (G0.31-0.20) seems to have a diffuse emission component which envelops the central, more compact core. These structures are also evident in the 1 GHz data of \citet{Heywood2022} and at far-IR wavelengths such as in the 70 $\rm\mu$m Herschel PACS data presented in \citet{Molinari2011}.

\subsection{SNTF2 total-intensity}
SNTF2 (G0.15-0.20) is located directly south of the portion of the Radio Arc observed in \citet{Pare2019} and \citet{Pare2021} and is the southern extension of the Radio Arc, as can be seen in Figure \ref{fig:chart}. This portion of the Radio Arc has previously been investigated \citep{YM1987}; however, the new VLA observations presented here provide an opportunity to conduct a more detailed study of this region. Furthermore, the VLA observations presented in this work extend further south than those shown in \citet{YM1987}.

\begin{figure*}
    \centering
    \plotone{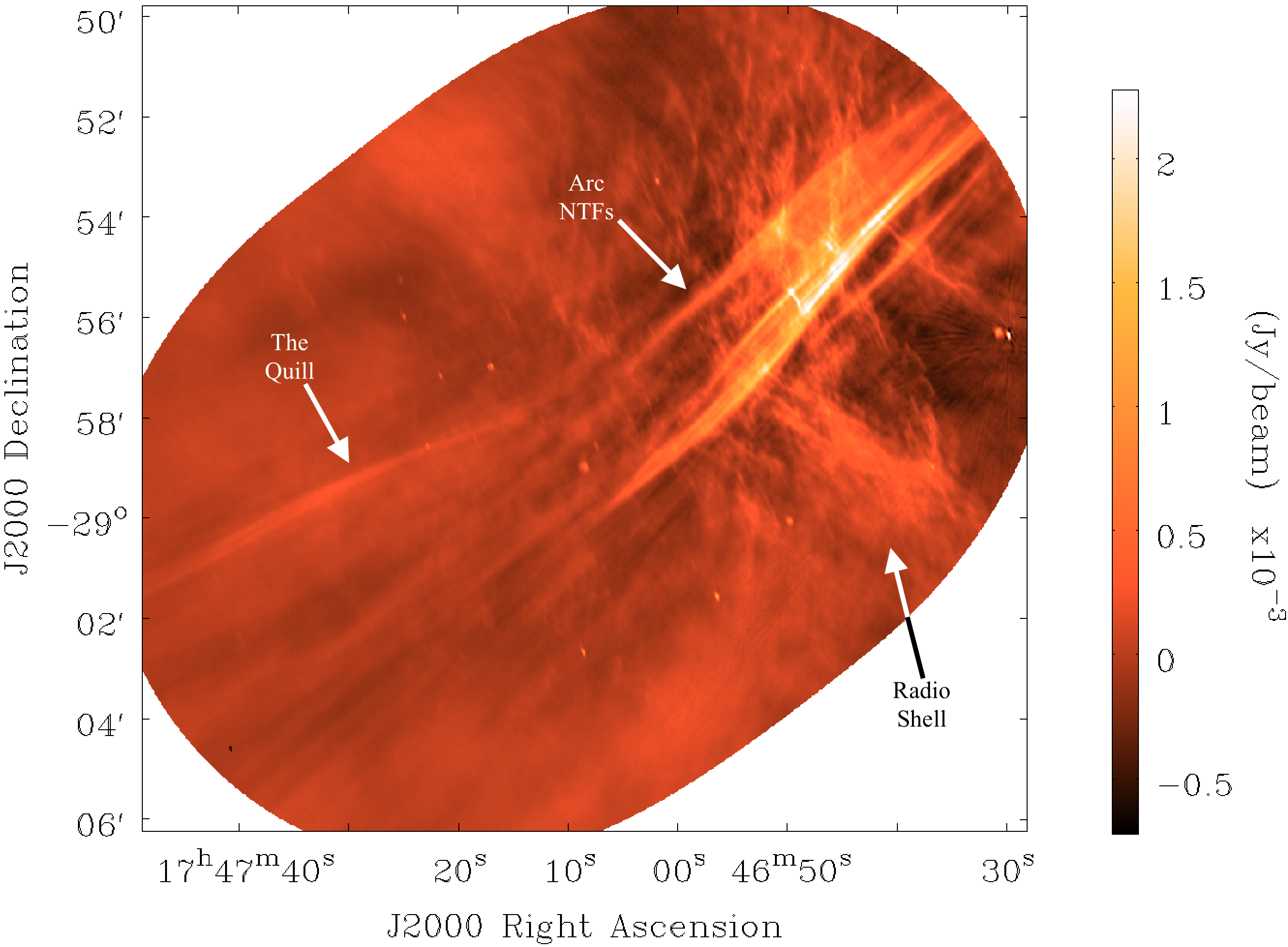}
    \caption{6 GHz total-intensity distribution of SNTF2 (G0.15-0.20). This image has a beam size of 4.7$\rm\arcsec$ x 2.1$\rm\arcsec$ and an rms noise level of $\rm\sim{}200\mu$Jy beam$\rm^{-1}$. Key features are labelled in the figure.}
    \label{fig:SNTF2_C}
\end{figure*}
\begin{figure*}
    \centering
    \plotone{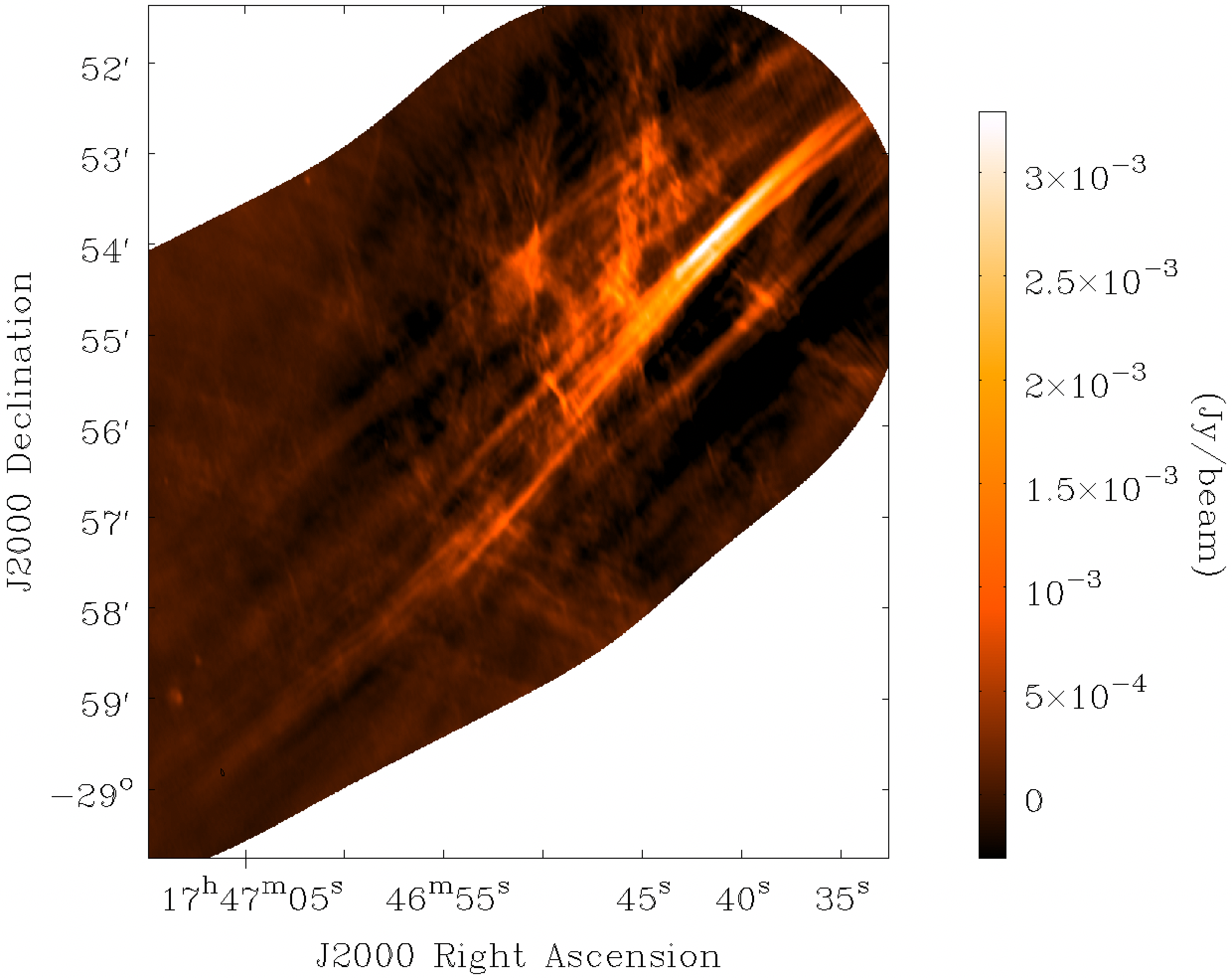}
    \caption{10 GHz total-intensity distribution of SNTF2 (G0.15-0.20). This image has a beam size of 4.6\arcsec\ x 2.1\arcsec\ and an rms noise level of $\rm\sim{}20\mu$Jy beam$\rm^{-1}$.}
    \label{fig:SNTF2_X}
\end{figure*}
\begin{figure}
    \centering
    \plotone{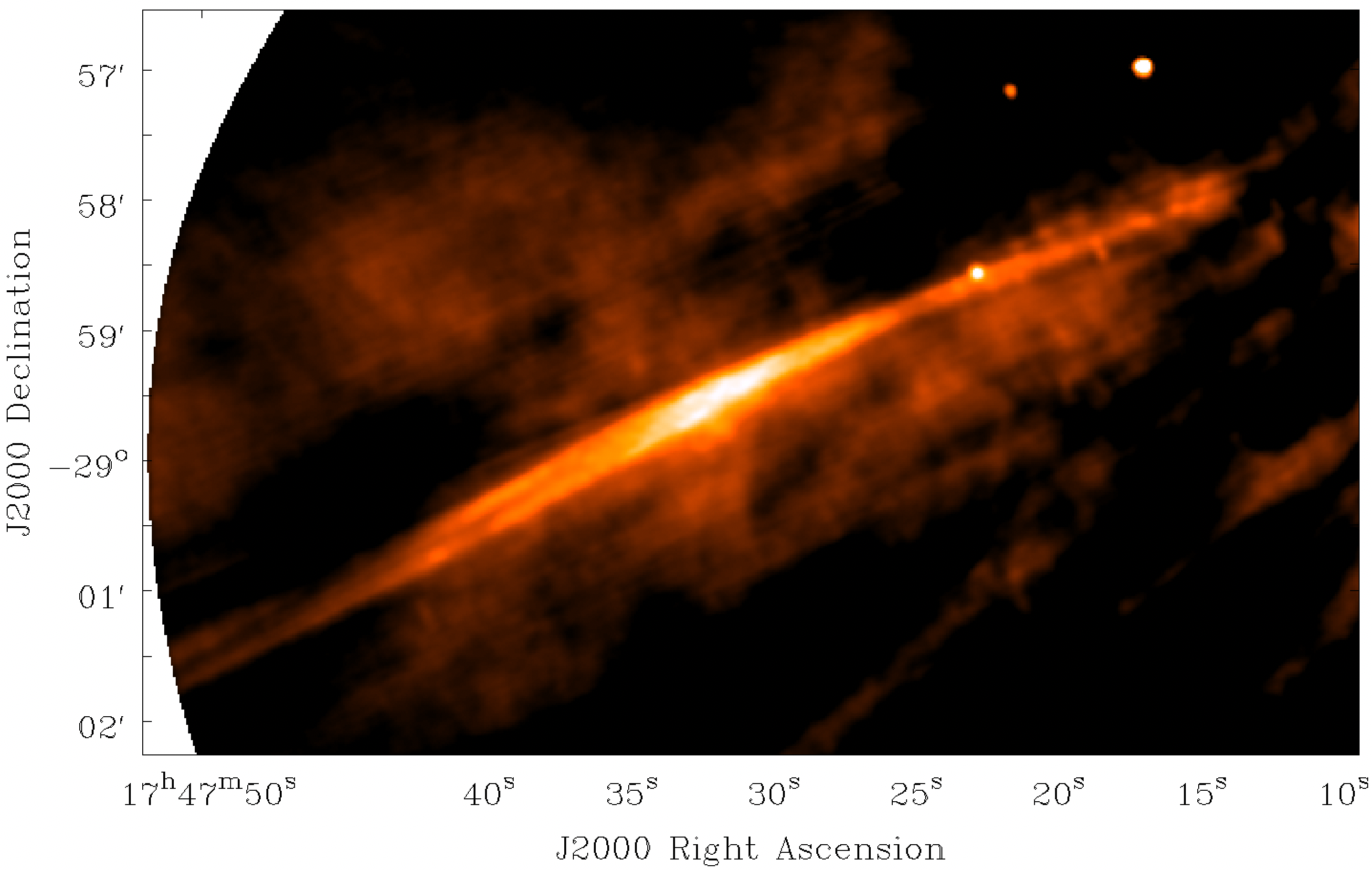}
    \caption{Zoom--in view of `The Quill' Filament (G0.17-0.39) at 6 GHz. Beam size and rms noise level are the same as in Figure \ref{fig:SNTF2_C}. The contrast in this image has been adjusted to emphasize the emission from The Quill.}
    \label{fig:Southern_I}
\end{figure}
\begin{figure}
    \centering
    \plotone{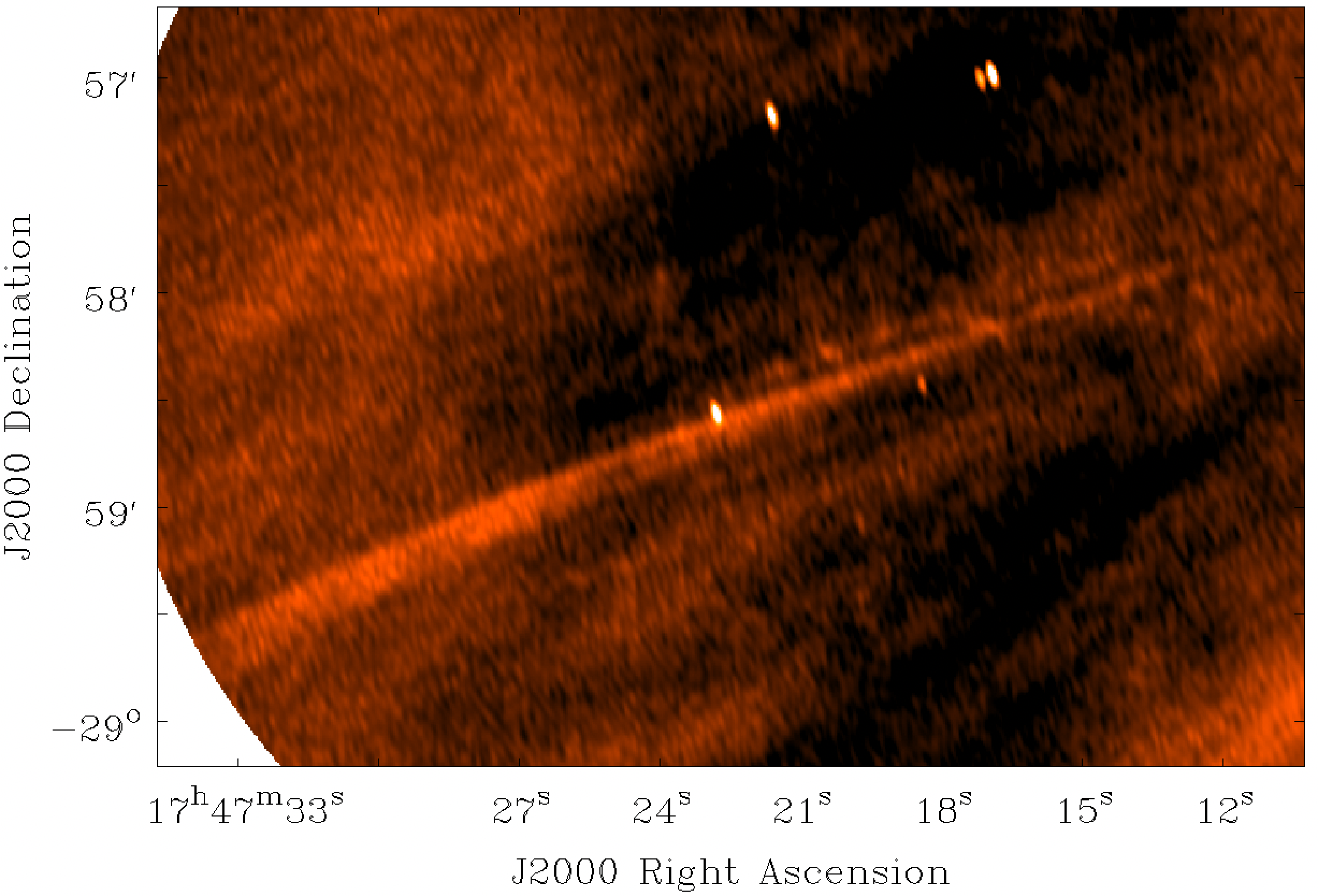}
    \caption{Zoom--in view of `The Quill' filament (G0.17-0.39) at 10 GHz. Beam size and rms noise level are the same as in Figure \ref{fig:SNTF2_X}. The contrast in this image has been adjusted to emphasize the emission from the Quill.}
    \label{fig:Quill_X}
\end{figure}
Figure \ref{fig:SNTF2_C} shows our 6 GHz intensity distribution for SNTF2, revealing that it is comprised of numerous individual filaments. The filaments of SNTF2 are not evenly spaced throughout the width of the structure, but rather appear to concentrate into discrete clumps with each clump consisting of multiple sub-filaments. Multiple point sources can also be identified throughout the image. 

Figure \ref{fig:SNTF2_X} in turn reveals the total-intensity distribution of SNTF2 at a central frequency of 10 GHz. We see the same morphology in our 10 GHz data set that we observe at 6 GHz. We cannot determine the total filament length from the observations presented in this work since we only observe the southern portion of the Radio Arc, but previous observations have found filament lengths of up to 45$\rm\arcmin$ (110 pc,\citealt{Yusef-Zadeh1986a,Pare2019}). We observe an average width of 4.0$\rm\arcsec$ (0.16 pc) at 10 GHz for the individual filaments comprising SNTF2. Using a quadrature decomposition and a beam width of 2.5\arcsec\ we find an intrinsic width of 3.1\arcsec\ (0.12 pc). The observed and intrinsic widths of the filaments in SNTF2 are shown in Table 2.

There are other extended total-intensity structures seen near SNTF2, that are roughly perpendicular to the SNTF2 filaments. These structures are labelled `Radio Shell' in Figure \ref{fig:SNTF2_C}. They are more easily seen at 6 GHz and are likely to be portions of the Radio Shell identified in previous observations of the Radio Arc (e.g. \citealt{Simpson2007,Pare2019,Pare2021}).

The brightness of the individual filaments comprising SNTF2 is generally constant along their lengths; however, there is a brightness enhancement at the region where the filaments of SNTF2 coincide with the observed portion of the Radio Shell, as seen in Figure \ref{fig:SNTF2_C}. This brightness enhancement is also observed at 10 GHz as seen in Figure \ref{fig:SNTF2_X} and it was observed in the previous VLA 6 GHz observations of this region of the Radio Arc presented in \citet{YM1987}. This brightness enhancement is brighter (2.4 mJy beam$\rm^{-1}$ at RA = 17.46.38, DEC = -29.55.5) than the combined brightnesses of the filaments in SNTF2 (1.0 mJy beam$\rm^{-1}$) and the Radio Shell (0.8 mJy beam$\rm^{-1}$). The elevated brightness of this region of the NTFB could indicate that the NTFB and Radio Shell are interacting.

There is also a more isolated NTF labelled `The Quill'  located at 17:47:32, -28:59:30 (G0.17-0.39) which seems to be separate from SNTF2. This NTF is fainter than the filaments comprising SNTF2. The Quill bifurcates into two or more separate filaments at either end of the structure as seen in a zoomed-in view of the Quill at 6 GHz shown in Figure \ref{fig:Southern_I}. The Quill bulges in the central region of the structure and is also brighter than at its ends where it bifurcates. This brightness enhancement could be obscuring that the Quill consists of multiple filaments throughout its length. Indeed, at 10 GHz the Quill is observed to consist of 2 or more filaments throughout the observed length, as seen in Figure \ref{fig:Quill_X}. However, the 10 GHz data do not include the Quill region along the Quill where the structure attains maximal brightness.

The total length of the Quill as measured from the 6 GHz distribution is 7.0$\rm\arcmin$ (16 pc) with a width which varies from 4$\rm\arcsec$ (0.16 pc) to 20$\rm\arcsec$ (0.8 pc) as measured from the 10 GHz data. With quadrature decomposition and a beam width of 2.8\arcsec\ we find that the intrinsic width varies from 2.9\arcsec\ (0.11 pc )to 19.8\arcsec\ (0.77 pc). The average observed and intrinsic widths of the Quill are shown in Table 2. 

The Quill is also observed in the 1 GHz data shown in \citet{Heywood2022} and \citet{Yusef-Zadeh2022} and is labeled feature S5 in their papers. Intriguingly, the Quill is coincident with an X-ray filament detected using both XMM-Newton \citep{Ponti2015} and Chandra \citep{Wang2021}. The spatial coincidence of this X-ray emission and the radio emission detected in this work and in the MeerKAT 1 GHz data of \citet{Heywood2022,Yusef-Zadeh2022} strongly indicate that this emission is coming from the same structure. This possibility is further strengthened by the fact that the X-ray and radio emission from this structure have similar orientations. These ideas will be further explored and presented in a subsequent paper devoted to the nature of the Quill.

\subsection{SNTF3 total-intensity}
\begin{figure*}
    \centering
    \plotone{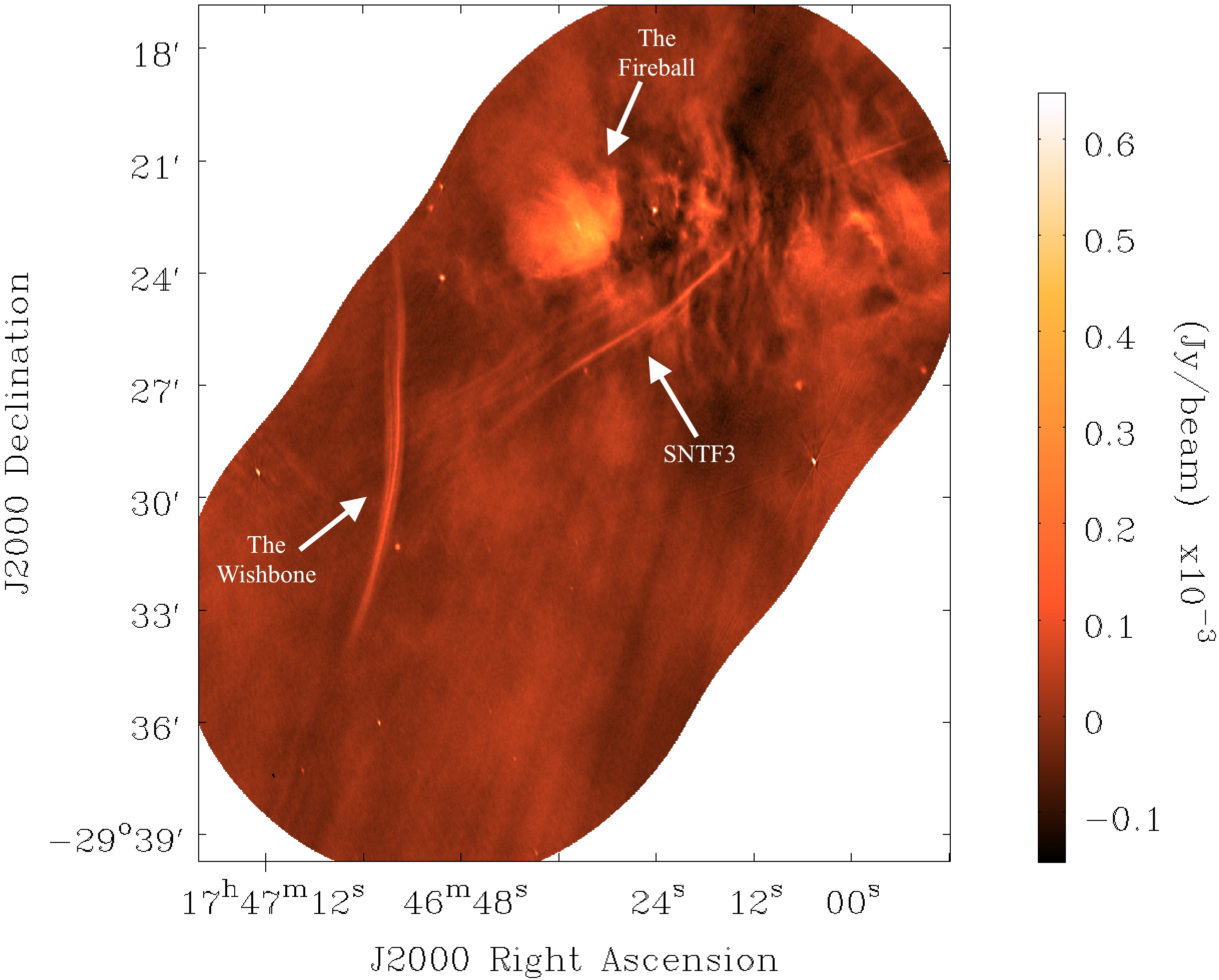}
    \caption{6 GHz total-intensity distribution of SNTF3 (G359.69-0.41). This image has a beam size of 4.5$\rm\arcsec$ x 2.0$\rm\arcsec$ with an rms noise level of $\rm\sim{}42\mu$Jy beam$\rm^{-1}$. Key features discussed in the text are labelled in the figure.}
    \label{fig:SNTF3_C}
\end{figure*}

SNTF3 (G359.69-0.41) is located about half a degree from Sgr A$\rm^*$ ($\rm\sim$70 pc away). It is located well below the Galactic Plane, as seen in Figure \ref{fig:chart}. In fact, SNTF3 is perhaps the most southerly Galactic NTF structure observed in the GC to date. The 6 GHz intensity distribution for SNTF3 is shown in Figure \ref{fig:SNTF3_C}. In addition to the SNTF3 structure, an additional NTFB is detected at 6 GHz (the Wishbone) and a number of prominent point sources can be observed.

\begin{figure*}
    \centering
    \plotone{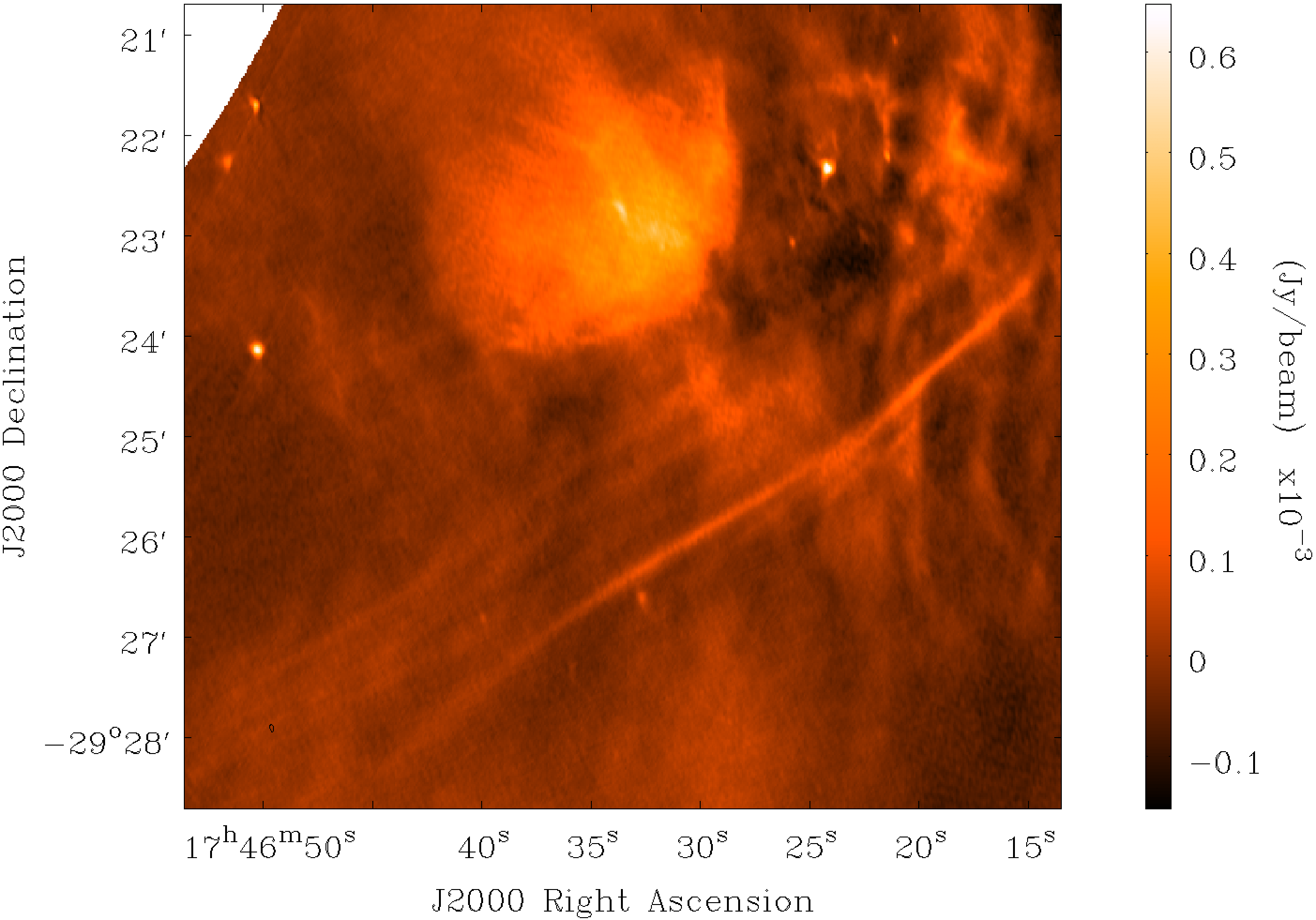}
    \caption{Zoom--in 6 GHz view of SNTF3 (G359.69-0.41). Beam size and rms noise level are the same as in Figure \ref{fig:SNTF3_C}.}
    \label{fig:SNTF3_Zoom_I}
\end{figure*}
\begin{figure}
    \centering
    \plotone{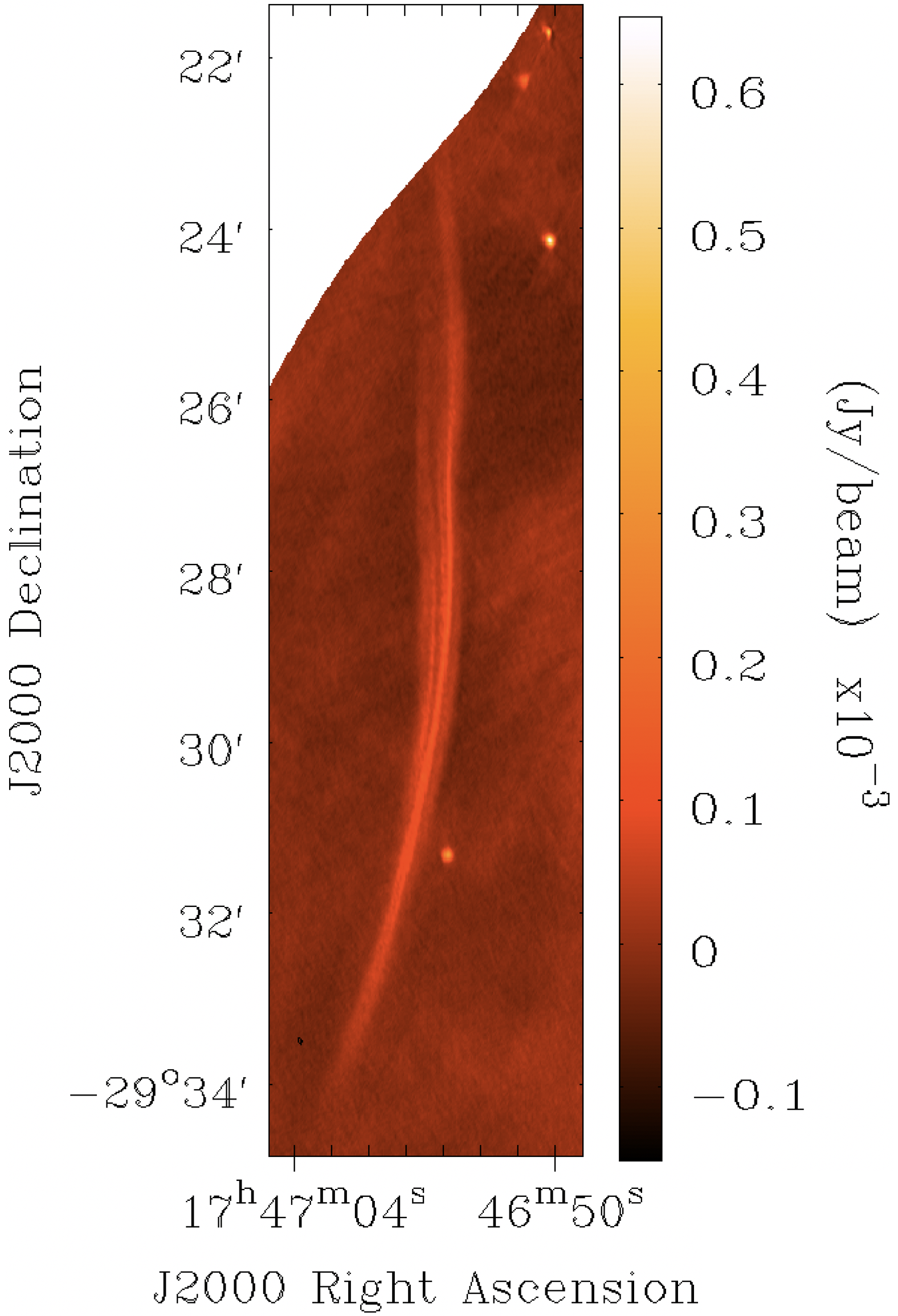}
    \caption{Zoom--in 6 GHz view of The Wishbone (G359.69-0.52). Beam size and rms noise level are the same as in Figure \ref{fig:SNTF3_C}.}
    \label{fig:wishbone_I}
\end{figure}
\begin{figure*}
    \centering
    \plotone{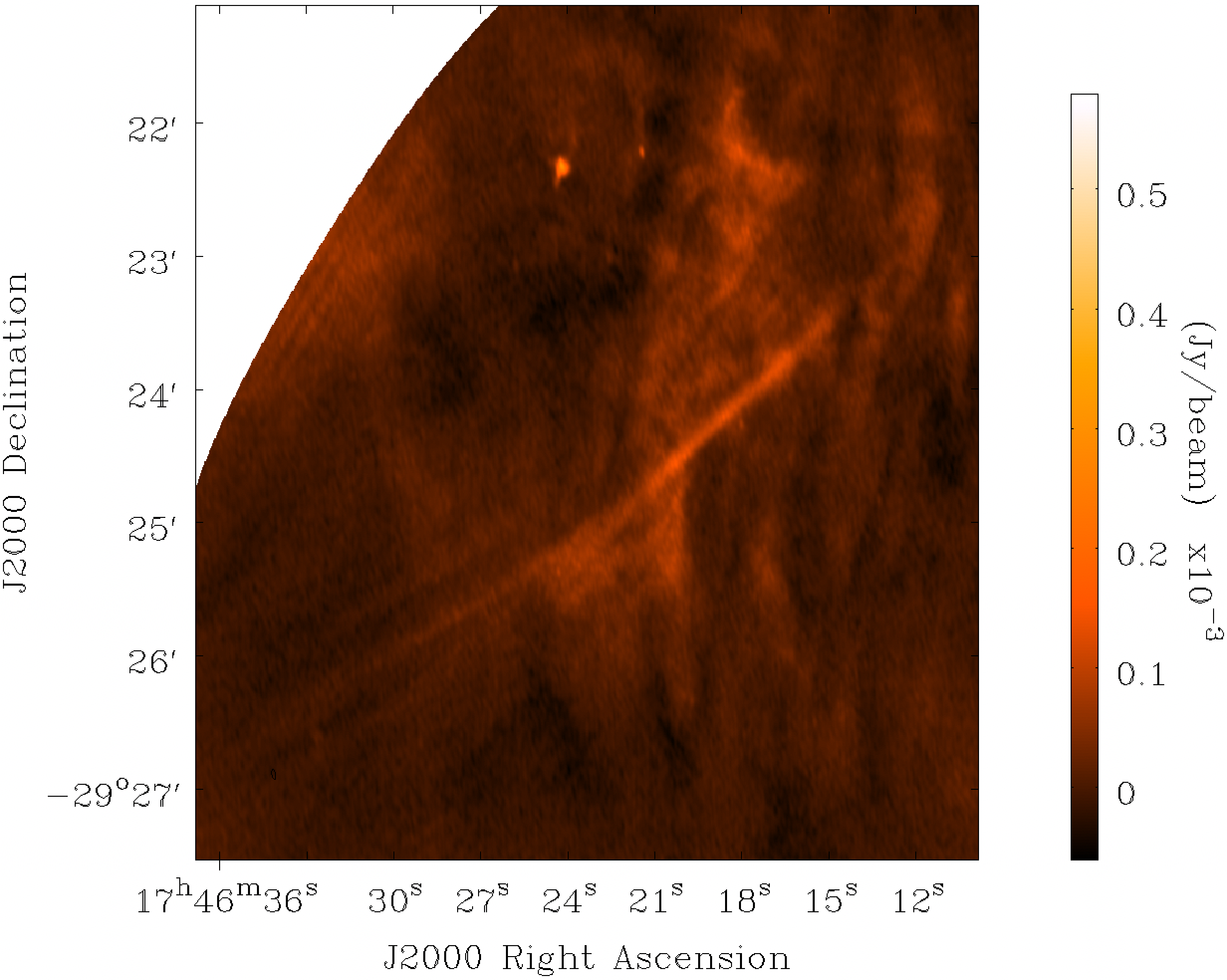}
    \caption{10 GHz total-intensity distribution of SNTF3 (G359.69-0.14). This image has a beam size of 4.6\arcsec\ x 1.7\arcsec\ with an rms noise level of $\rm\sim{}20\mu$Jy beam$\rm^{-1}$.}
    \label{fig:SNTF3_X}
\end{figure*}
SNTF3 is comprised of a cluster of faint filaments, as shown in the northern portion of Figure \ref{fig:SNTF3_C}. One of the filaments in this structure is much brighter than the others. A zoomed-in view of SNTF3 is shown in Figure \ref{fig:SNTF3_Zoom_I}. This view reveals the fainter filaments in SNTF3 in more detail. Unlike SNTF1 and SNTF2, the filaments of SNTF3 are evenly spaced and remain strictly parallel to one another. There are also no notable brightness enhancements within the filaments of SNTF3. SNTF3 has also been observed at 1 GHz and is shown to possess the same morphology as observed in this work (\citealt{Heywood2022,Yusef-Zadeh2022} their feature A19).

In addition, the 6 GHz image shows an isolated bundle of tightly packed filaments located at 17:46:56, -29:28:29 (G359.69-0.52), labelled `The Wishbone' in Figure \ref{fig:SNTF3_C}. A zoom-in image of The Wishbone is shown in Figure \ref{fig:wishbone_I}. This NTFB has a roughly 60 degree angle compared to the orientation of the SNTF3 cluster. This different orientation indicates that the Wishbone is likely not associated with SNTF3 and is therefore a distinct structure. 

The Wishbone consists of multiple filaments along most of its length. At its southern-most extent, however, the filaments seem to converge into a single filament as seen in Figure \ref{fig:wishbone_I}. The wishbone has a length of 9.4$\rm\arcmin$ (22 pc) from end-to-end, with a width that varies from 5$\rm\arcsec$ (0.2 pc) to 15$\rm\arcsec$ (0.6 pc) as measured from our 6 GHz data, where the thicker portions of the Wishbone conincide with the region where the filaments seem to converge. Using quadrature decomposition and a beam width of 2.1\arcsec\ we find an intrinsic width which varies from 4.5\arcsec\ (0.17 pc) to 14.9\arcsec\ (0.58 pc). The average observed and intrinsic widths of a filament within the Wishbone are also shown in Table 2. The portion of the Wishbone that appears as a single filament is generally thicker than the rest of the NTF, which could be a result of the superposition of the sub-filaments along the line of sight or an actual splitting of the broad filament into multiple sub-filaments. The Wishbone is also observed in \citet{Heywood2022} and \citet{Yusef-Zadeh2022} at 1 GHz, and is labelled as feature A15 in their papers.

There is also a diffuse structure located to the north of SNTF3 at 17:46:33.4, -29.22.56.8 (G359.73-0.41), labelled `The Fireball' in Figure \ref{fig:SNTF3_C}. There is a compact region of enhanced intensity at the center of the structure, and there is falloff from there toward the edge of the structure as can be seen in Figure \ref{fig:SNTF3_Zoom_I}. This structure is detected at both 1 GHz and 70 $\rm\mu$m as well \citep{Heywood2022,Yusef-Zadeh2022,Molinari2011}. This structure also coincides with an enhancement of atomic hydrogen as seen in Figure 4 of \citet{Molinari2011}.

The 10 GHz total-intensity distribution of SNTF3 is shown in Figure \ref{fig:SNTF3_X}. At these higher frequencies, because of the smaller field of view, the Wishbone NTF is not visible. In addition, the diffuse structure labeled as the Fireball in Figure \ref{fig:SNTF3_C} is at the edge of the field of view. At 10 GHz we only detect the brightest filament of SNTF3. The overall length of this filament in SNTF3 is 4.3$\rm\arcmin$ (10 pc), with an average width of 4.5$\rm\arcsec$ (0.17 pc) as measured using the 10 GHz data shown in Figure \ref{fig:SNTF3_X}. Using quadrature decomposition and a beam width of 3.2\arcsec\ we find the average intrinsic width to be 3.2\arcsec\ (0.12 pc). The average observed and intrinsic widths of SNTF3 are shown in Table 2.

\subsection{Spectral Index Values} \label{sec:spec_val}
\begin{figure*}
    \centering
    \gridline{\fig{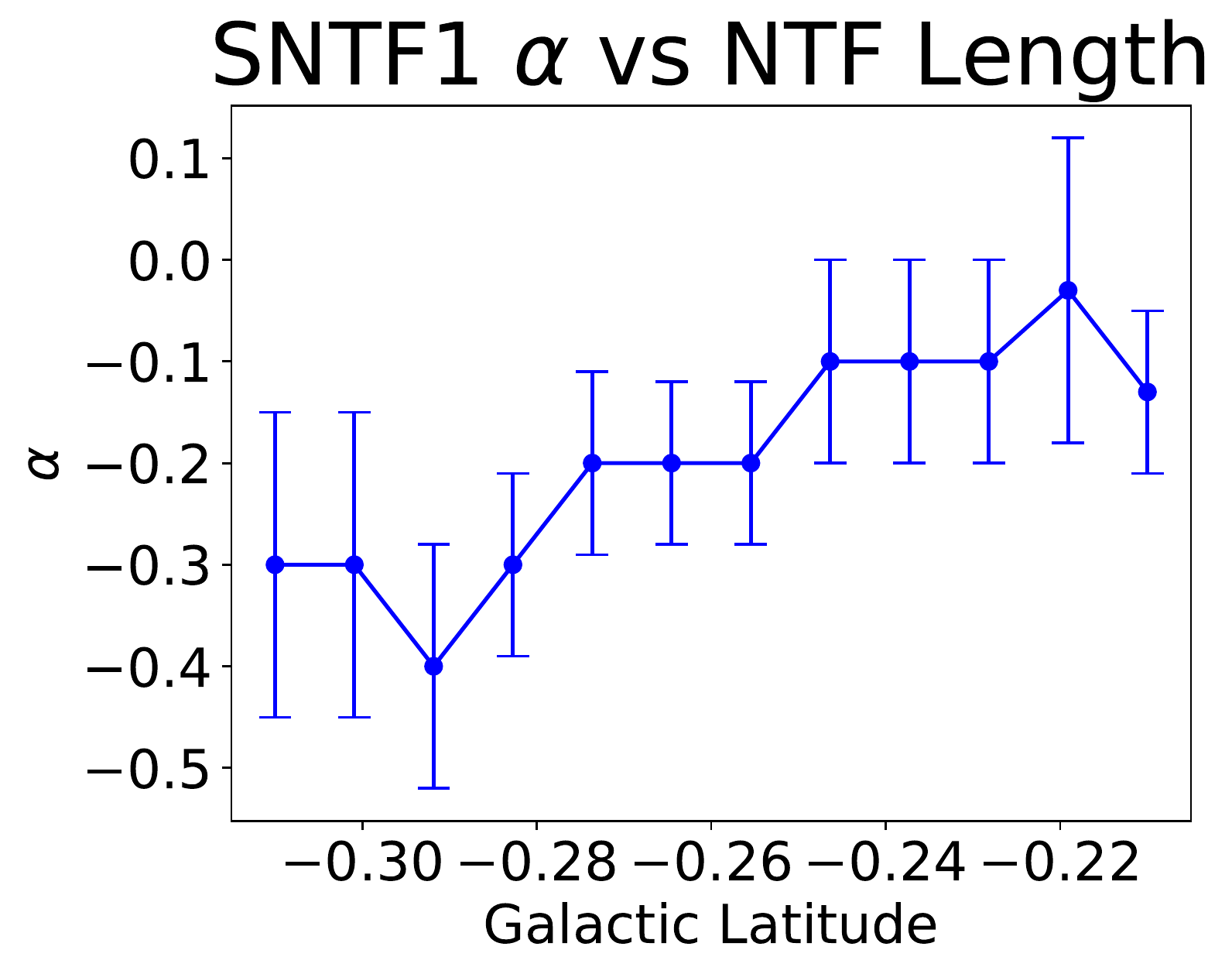}{0.3\textwidth}{a}
              \fig{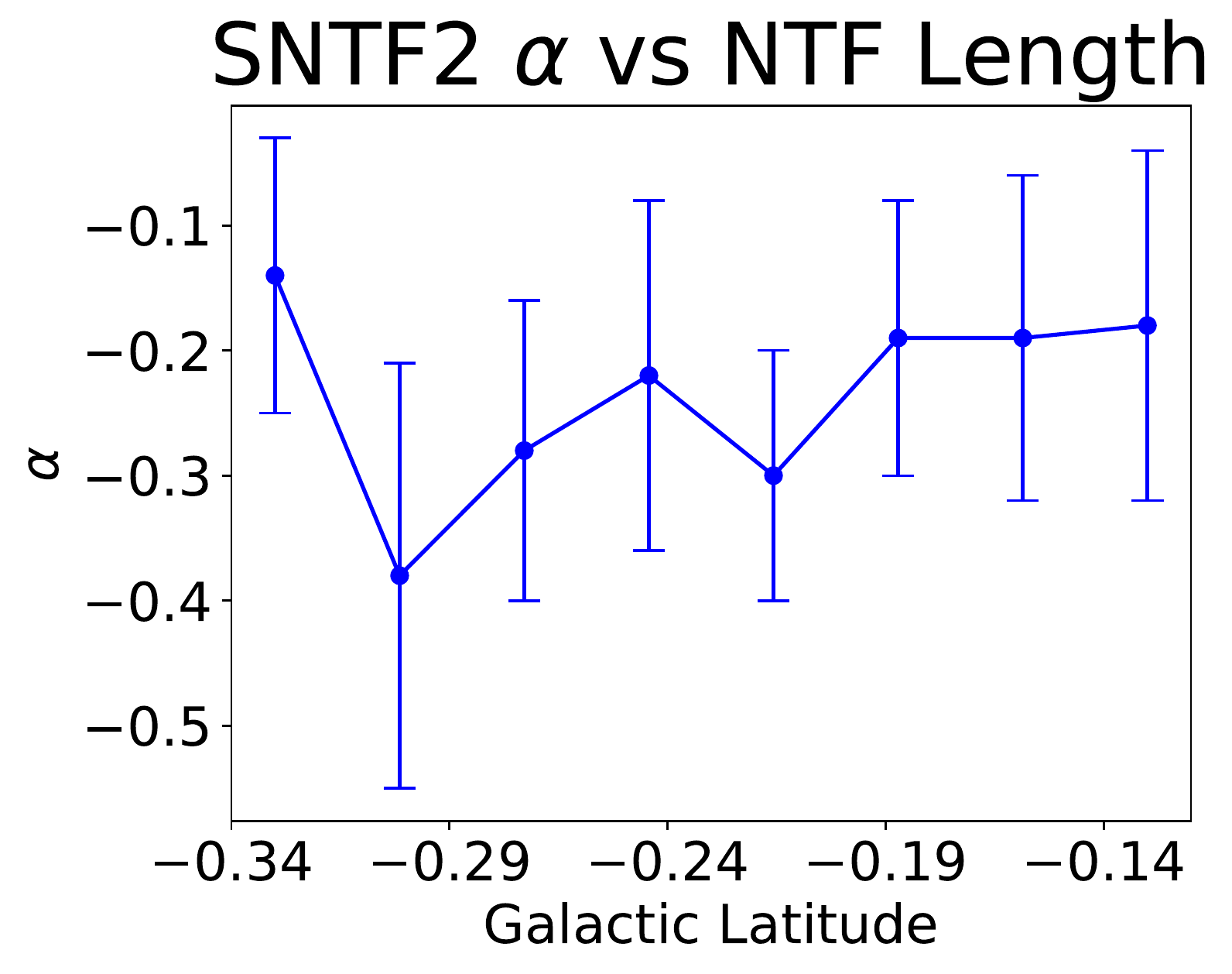}{0.3\textwidth}{b}
              \fig{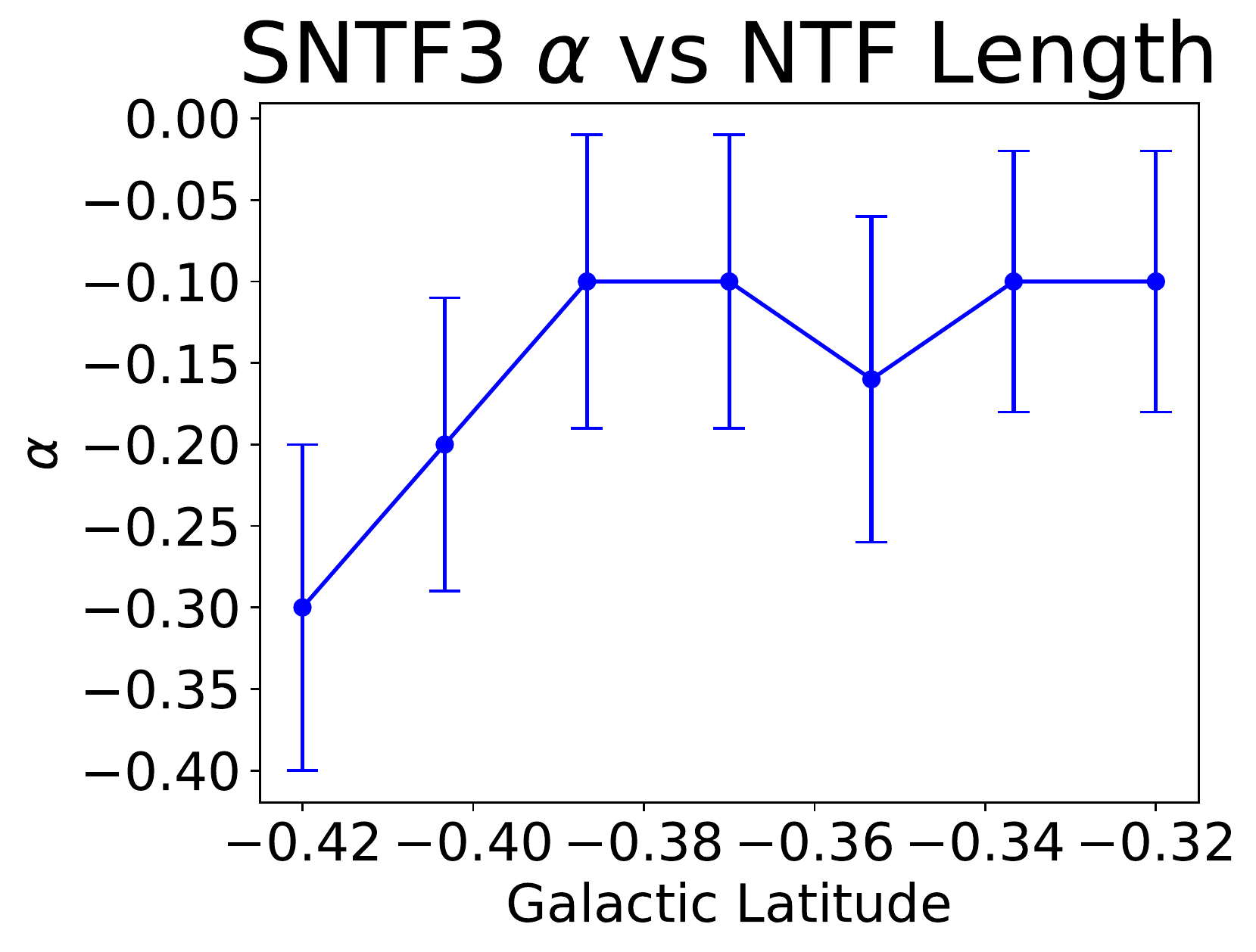}{0.3\textwidth}{c}}
    \gridline{\fig{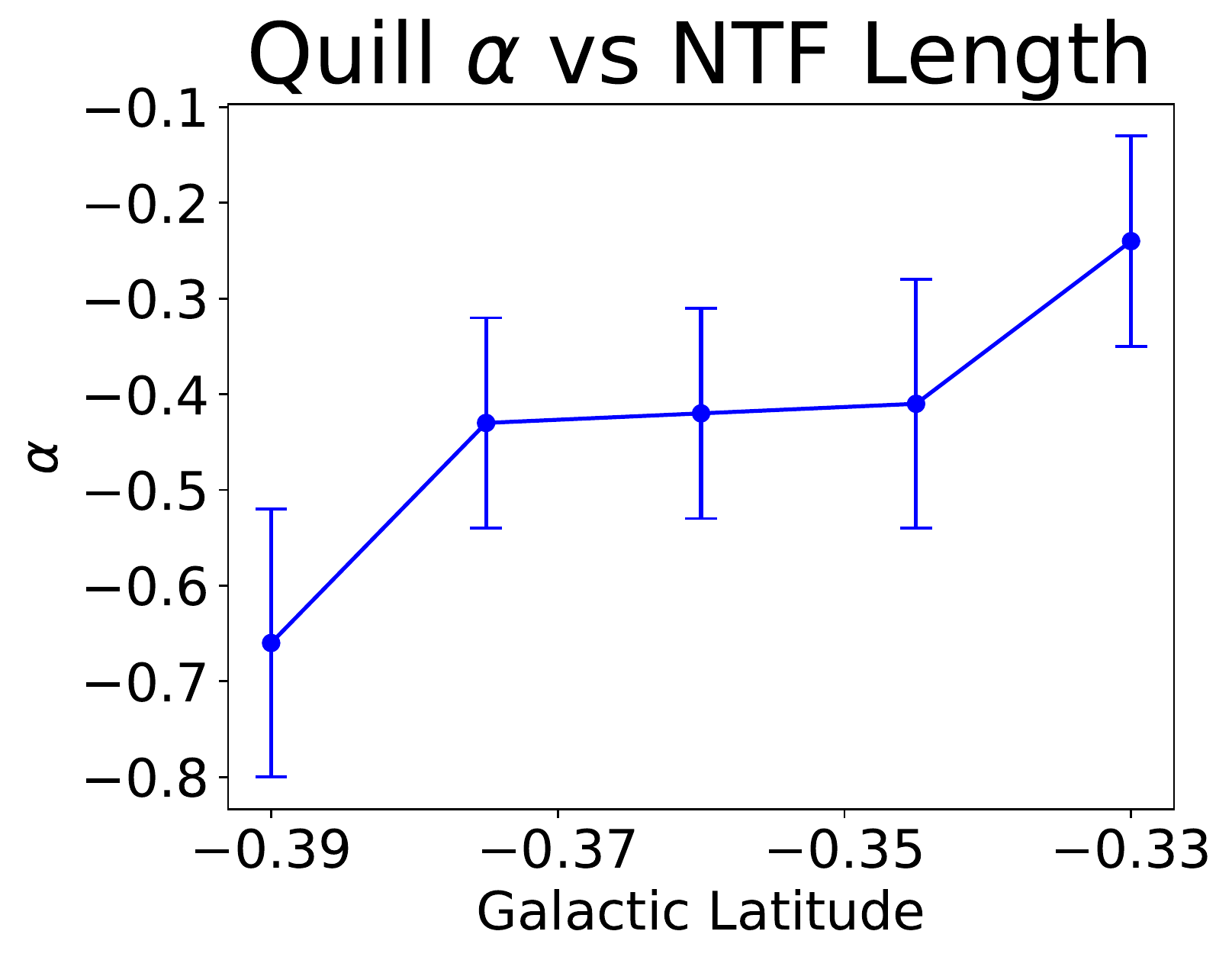}{0.3\textwidth}{d}
              \fig{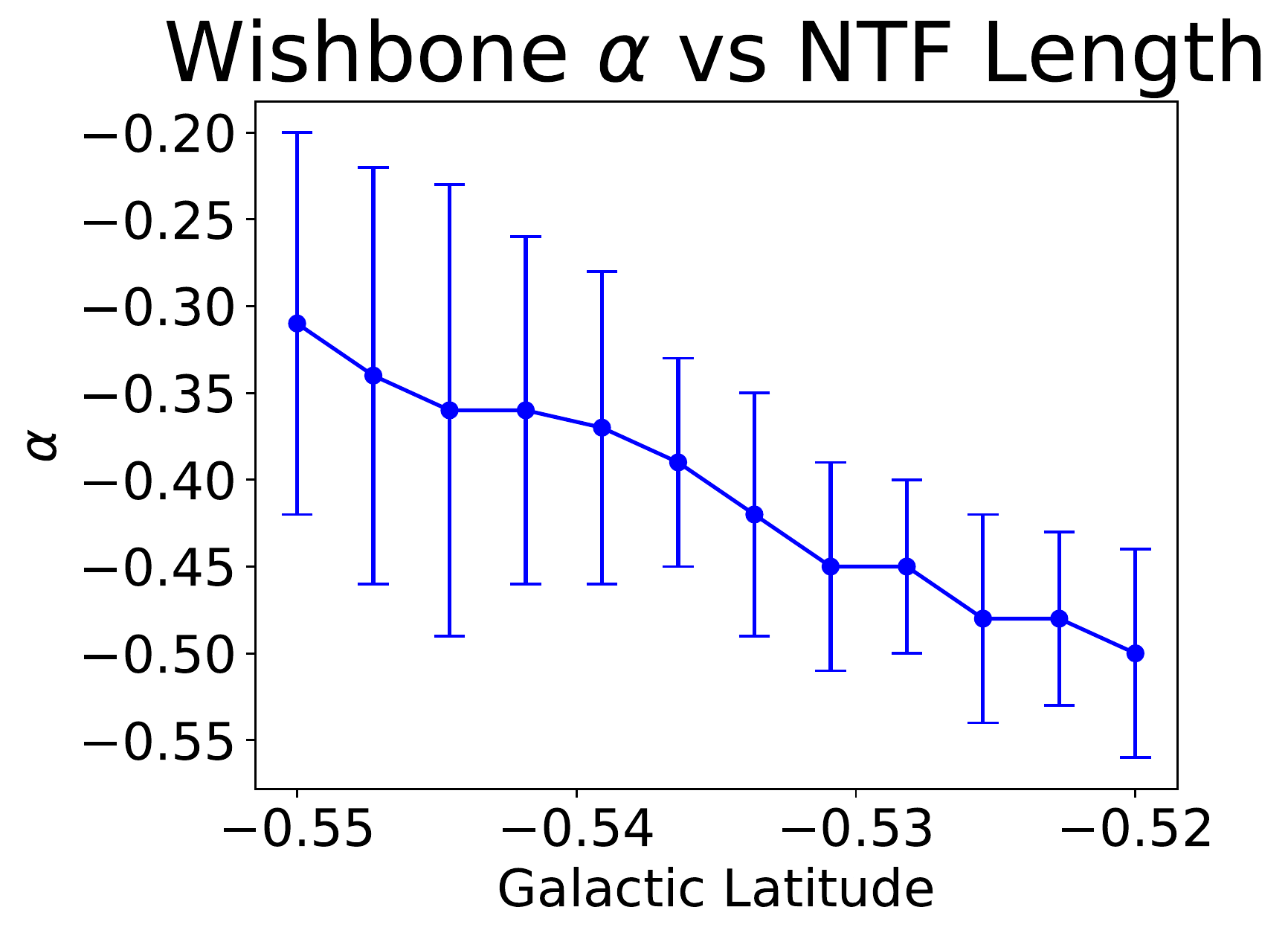}{0.3\textwidth}{e}}
    \caption{Spectral index as a function of Galactic latitude averaged over 20\arcsec\ segments of the NTFs. The error bars represent the standard deviation of the spectral index within each of these 20\arcsec\ segments. The change in latitude from left to right in these panels roughly corresponds to tracing the filament lengths from the lower left to the upper right in the figures presented in Section \ref{sec:res}.}
    \label{fig:NTF_alpha}
\end{figure*}
We derive the spectral indices of our targets using the method described in Section \ref{sec:a_deriv}. The spectral indices as a function of Galactic latitude for our targets are shown in Figure \ref{fig:NTF_alpha}. The change in latitude corresponds to tracing the NTFs from the lower left to the upper right (from higher RA and lower DEC to lower RA and higher DEC) of the total-intensity images presented in Section \ref{sec:res}. The error bars are the standard deviations of the spectral indices within each 20\arcsec\ segment.

SNTF1 has an average spectral index of -0.2 $\rm\pm$0.1 as can be seen in panel a of Figure \ref{fig:NTF_alpha}. The spectral index for SNTF2 (panel b) reveals an average spectral index of -0.2 $\rm\pm$0.1. SNTF3 has an average spectral index of -0.1 $\rm\pm$0.1 (panel c). Both the Quill and Wishbone have average spectral indices of -0.4 $\rm\pm$0.1 (panels d and e of Figure \ref{fig:NTF_alpha}, respectively). We find negative spectral indices throughout the lengths of our NTFBs. the average spectral indices for the NTFBs are shown in Table 2. 

Though there is no convincing evidence of spectral index gradients in the majority of our targets, there is a clear gradient in the spectral index of the Wishbone. The Wishbone has a systematically shallower spectral index of -0.3 at lower latitudes and a steeper spectral index of -0.5 at higher latitudes as can be seen in panel e of Figure \ref{fig:NTF_alpha}.

\subsection{Equipartition Estimates} \label{sec:equi}
\begin{deluxetable*}{|c|c|c|c|c|c|c|c|c|c|}
\tablecaption{Key Properties of Observed NTFs}
\tablecolumns{10}
\tablenum{2}
\tablewidth{0pt}
\tablehead{
\colhead{Obj.} & \colhead{Coord.} & \colhead{$\rm\alpha_{1 GHz}$} & \colhead{$\rm\alpha_{6 - 10 GHz}$} & \colhead{$\rm\alpha$ Grad.} & \colhead{L} & \colhead{W} & \colhead{IW} & \colhead{$\rm{}S_{\nu}$} & \colhead{$\rm{}B_{eq}$}
}
\startdata
SNTF1 & G0.30-0.26 & -0.5 & -0.2$\rm\pm$0.1 & No & 15 & 0.12 & 0.09 & 1.2 & 90 \\
SNTF2 (Radio Arc) & G0.15-0.20 & -0.2 & -0.2$\rm\pm$0.1 & No & 110 & 0.16 & 0.12 & 3.5 & 600 \\
SNTF3 & G359.69-0.41 & -0.8 & -0.1$\rm\pm$0.1 & No & 10 & 0.17 & 0.12 & 0.2 & 300 \\
Quill & G0.17-0.39 & -0.5 & -0.4$\rm\pm$0.1 & No & 16 & 0.16 & 0.11 & 0.1 & 200 \\
Wishbone & G359.69-0.52 & -0.5 & -0.4$\rm\pm$0.1 & Yes & 22 & 0.2 & 0.17 & 0.15 & 300
\enddata
\tablecomments{Obj. Name indicates the name of the NTF, Coord. indicates the Galactic coordinates of the object, $\rm\alpha_{1 Ghz}$ is the spectral index observed in \citet{Yusef-Zadeh2022}, $\rm\alpha_{6 - 10 GHz}$ is the average spectral index seen for the NTFB in this work as derived from the spectral indices shown in Figure \ref{fig:NTF_alpha}, $\rm\alpha$ Gradient indicates whether the NTFB is observed to have a spectral index gradient based on the data presented in this work, L is the length of a filament within the NTFB in pc, W is the observed width of a filament within the NTFB in pc, IW is the intrinsic width of a filament within the NTFB in pc as determined from quadrature decomposition, $\rm{}S_{\nu}$ is the flux density of the source in mJy beam$\rm^{-1}$ s$\rm^{-1}$ Hz$\rm^{-1}$, and $\rm{}B_{eq}$ is the equipartition magnetic field strength estimated for the NTF in $\rm\mu$G.}
\end{deluxetable*}
Following in the steps of previous NTF studies we can estimate the minimum magnetic field strengths local to the NTFBs. One way to make this estimate is by assuming the magnetic energy density is in equipartition with the kinetic energy of relativistic particles within the NTFBs. This assumption can be applied in a region where energy exchange takes place between particles and fields on time scales smaller than the energy loss time of the particles or the field generation time from particle dynamics. This assumption therefore may not accurately characterize magnetic field strength in the GC region, given the large-scale order of the magnetic field \citep{Morris2006}. 

However, an equipartition estimate can still provide insight into the minimum magnetic field strength of the NTFB targets of this work. To determine these minimum magnetic field strengths we follow the formalism of \citet{Moffet1975}, \citet{Dubner2008}, and \citet{Dubner2015}. The assumption of equipartition has been used to estimate the magnetic field strengths for numerous other NTFs in the GC (e.g. \citealt{Gray1995,Lang1999a,Lang1999b}). Since there is no obvious cutoff frequency for our targets, we use the standard lower and upper cutoff frequencies of 10$\rm^7$ Hz and 10$\rm^{11}$ Hz respectively. To determine the volume of our NTFBs we assume cylindrical symmetry and use the average intrinsic width of individual filaments within the NTFBs as determined from the 6 and 10 GHz data as the radius of the cylindrical volume. We use the average luminosity of the NTFBs as the characteristic luminosity throughout the NTFBs. The average luminosities for the NTFBs, as well as the parameters for the cylindrical volumes of the filaments within the NTFBs, are shown in Table 2.

We find minimum magnetic field estimates ranging from 90 - 300 $\rm\mu$G for the filaments comprising SNTF1, SNTF3, the Wishbone, and the Quill, as shown in Table 2. For the filaments in SNTF2, however, which is a segment of the Radio Arc, we find a significantly higher minimum magnetic field strength estimate of 600 $\rm\mu$G. This can be attributed to the fact that the Radio Arc is a brighter structure than our other NTFB targets, as seen in Table 2, and to the proportionality of the magnetic field strength to the radio continuum intensity under the equipartition assumption.

\section{DISCUSSION} \label{sec:disc}

\subsection{Comparison of the NTFBs with the Larger NTF Population}
One of the first questions to address regarding the targets studied in this work is whether they have similar properties to the larger NTF population. The total-intensity and morphologies of our targets are similar to what is observed in the larger NTF population. 

The lengths of the filaments within most of our targets are on the order of a few arcminutes ($\rm\sim$5.5\arcmin), which agrees with the lengths of most of the NTFs observed in the GC \citep{Lang1999a,Lang1999b,YWP1997}. The Radio Arc, which contains SNTF2, is much longer with filament lengths of 45\arcmin\ (110 pc), similar to the longest known NTFs like the Snake and C16, which have lengths of 24\arcmin\ and 40\arcmin\ (56 pc and 100 pc) respectively. Furthermore, the widths of individual filaments within our targets are $<<$1 pc, a property which is also observed in the larger NTF population \citep{Morris1996b,Yusef-Zadeh2004,Yusef-Zadeh2022}.

The brightness variations and bends observed in our NTFBs seen in Section \ref{sec:res} are also observed in multiple other NTFs. The Snake, for example, exhibits both brightness enhancements and kinks along its length \citep{Gray1995}. The bending observed in all 5 of our NTFB targets is also a commonly observed feature of the larger NTF population. The Pelican, the Northern Thread, and the Southern Thread, for example, all reveal bends which occur somewhere along the filament lengths \citep{Lang1999a,Lang1999b}.

We find that the spectral indices of our targets are negative throughout their lengths, which agrees with the spectral indices observed for the larger NTF population \citep{Yusef-Zadeh2022}. The spectral indices of our targets are generally shallower than what is observed in the larger NTF population, however. \citet{Yusef-Zadeh2022} observe an average spectral index of -0.8 for the larger NTF population, whereas we see an average spectral index of -0.3 for our NTFB targets. However, shallower spectral indices like those observed for our targets are not unprecedented. The Radio Arc has long been known to have a shallow (essentially flat) spectral index of $\rm\sim$\,-0.1 \citep{Mills1984,Yusef-Zadeh1987}, and our results corroborate this finding.

The total-intensity, morphology, and spectral index properties of our targets indicate that they are indeed members of the larger NTF population. We now explore how the different individual NTFBs studied in this work compare to other previously-studied NTFs.

\subsection{SNTF1 and SNTF3 Compared to the Radio Arc}
Comparing the widths of the SNTF bundles, we observe that SNTF1 and SNTF3 have similar total widths of 1.3\arcmin\ and 1.4\arcmin\ respectively (2.8 and 2.9 pc). In comparison, the Radio Arc has a total width of 2.4$\rm\arcmin$ (6.2 pc). Comparing the lengths and widths of SNTF1 and SNTF3 to the Radio Arc, we find that the Radio Arc is almost 20 times larger on the sky than SNTF1 and SNTF3. Therefore, though the individual filaments comprising all of the observed SNTFs exhibit similar properties, the Radio Arc is a significantly larger structure than either SNTF1 or SNTF3.

Since our observations cover the frequency range from 4 - 12 GHz, we can compare the spectral indices we derive over this frequency range with those obtained from the 1 GHz MeerKAT data presented in \citet{Yusef-Zadeh2022}. The spectral indices reported for our NTFB targets by \citet{Yusef-Zadeh2022} are also presented in Table 2. \citet{Yusef-Zadeh2022} observed  spectral index values for the Radio Arc (SNTF2) that are similar to what we derive from our higher frequency observations. However, we find a shallower spectral index for SNTF1 and SNTF3, -0.2, than what is derived from the 1 GHz MeerKAT data. 

We do not see convincing evidence of spectral index gradients in SNTF1, SNTF2, or SNTF3 (see panels a, b, and c of Figure \ref{fig:NTF_alpha}). \citet{Yusef-Zadeh2022} does not detect spectral index gradients for these targets from their 1 GHz data, and the majority of NTFs observed to date do not exhibit convincing spectral index gradients. The discrepancy in spectral index steepness for the SNTFs between our work and those reported by \citet{Yusef-Zadeh2022} could be a result of our inability to fully correct for the changing background in our GC observations. SNTF1 and SNTF3 are fainter than SNTF2 and would therefore be more affected by uncorrected background contamination.

Examining the locations of the SNTFs within the GC we see that both SNTF1 and the Radio Arc (SNTF2) extend to distances of $\rm\sim$0.25deg (35 pc) from the Galactic plane, whereas SNTF3 is almost twice as far from the Galactic plane at $\rm\sim$0.4deg (56 pc). Comparing the spatial locations of the observed SNTFs with their spectral indices our observations therefore tentatively indicate a trend of steeper spectral index closer to the Galactic Plane. This tentative trend from our observations does agree with the trend observed in \citet{Yusef-Zadeh2022} of flatter spectral indices at larger latitudes, which they claim could be a result of synchrotron cooling.

\subsection{Quill and Wishbone Compared to Larger NTF Population}
In addition to our SNTFs that are comprised of $\rm\geq$10 individual filaments, we also detect the Quill and Wishbone NTFBs that appear to be comprised of fewer filaments. The Quill and Wishbone therefore exhibit similar morphologies to the majority of previously-studied NTFs within the GC that are comprised of only an isolated filament or a few parallel filaments.

One possibility to consider for why our SNTF targets are comprised of a large number of individual filaments is the complexity of their local environments. The SNTFs observed in this work are close in projection to other objects like diffuse radio structures and other NTFs. However, we also observe the Quill and Wishbone projected toward the same relatively complex regions of the GC in which the Radio Arc (SNTF2) and SNTF3 are located. Furthermore, other NTFs throughout the GC are observed in locally complex GC regions, such as the Northern and Southern Threads \citep{Lang1999b}. Therefore, the complexity of the environment does not inherently dictate the number of filaments comprising an NTFB.

We find an average spectral index of -0.4 for both the Quill and the Wishbone. This spectral index is steeper than what is observed for our 3 SNTF targets, but in agreement with the spectral indices observed for the larger NTF population (e.g., \citealt{Yusef-Zadeh2022}). Our spectral index values for the Quill and Wishbone and those measured at 1 GHz in \citet{Yusef-Zadeh2022} agree within the 0.1 uncertainties of our measurements as shown in Table 2. Notably, we see convincing evidence of a spectral index gradient in the Wishbone NTFB where the spectral index varies smoothly from -0.30 at lower latitudes to -0.50 at higher latitudes as can be seen in panel e of Figure \ref{fig:NTF_alpha}. This gradient is not observed in the 1 GHz data presented by \citep{Yusef-Zadeh2022}.

The convincing gradient detected for the Wishbone could be a result of synchrotron energy losses as the relativistic electrons travel along the NTFB filaments from their acceleration sites \citep{Serabyn1994}. The filaments in the Wishbone are at their brightest at the most southerly latitudes where the spectral index is shallowest. This also coincides with the region where the individual filaments in the Wishbone seem to converge as seen in Figure \ref{fig:wishbone_I}. This southern portion of the Wishbone could therefore be closer to the source of relativistic electrons illuminating the NTFB.

\subsection{A Possible Formation Scenario for the NTFs} \label{sec:form}
A recent study by \citet{Thomas2020} used the 1 GHz MeerKAT data of the GC to observe a pair of NTFBs comprised of multiple parallel filaments (G359.85+0.39 and  G359.47+0.12) which they designate as `harps.' The two harps studied by \citet{Thomas2020} exhibit a systematic change in filament length from one side of the NTFB system to the other. \citet{Thomas2020} developed a theory to explain the formation of the harp NTFB structures, proposing that an object like a pulsar wind nebula could be producing relativistic electrons with energies of a few GeV which travel toward wind termination shocks. The magnetic field piling up behind this shock reconnects with the background magnetic field \citep{Thomas2019}. The high-energy electrons then diffuse along the reconnecting magnetic field lines, illuminating the NTFBs observed.  

There is ample evidence of elevated large-scale high-energy activity in the GC, such as regions of enhanced synchrotron emissivity \citep{Yusef-Zadeh2004,Yusef-Zadeh2007} and large-scale X-ray structures coincident with outflow structures identified in infrared and radio wavelengths  \citep{Ponti2019,Ponti2021}. The elevated magnetic field strengths observed in the GC \citep{Morris2006} also support the idea of an elevated CR population \citep{Yusef-Zadeh2007}. The CRs could have originated from the large-scale X-ray emission identified with the outflowing structures \citep{Ponti2021}. Alternatively, the CRs could have leaked from the NTFs which are seen to pervade the GC as can be seen in Figure \ref{fig:chart}.

In our NTFB targets we see a similar morphology for SNTF1 to what is observed for the harp structures in \citet{Thomas2020}, where the filaments are systematically longer on one side of the NTFB than the other. The side with the shortest filaments in SNTF1 is closer to the jet-like source, as can be seen in Figures \ref{fig:SNTF1_C}, \ref{fig:SNTF1_Zoom_C}, and \ref{fig:SNTF1_X}. The shorter filaments closer to the point source are also generally brighter than the longer filaments; therefore, the morphology of SNTF1 agrees with the model of cosmic ray transport developed in \citet{Thomas2020} for the harps. Indeed, SNTF1 can likely be categorized as another harp structure. The shorter filaments are likely shorter and brighter because they are the ones which have most recently been energized with relativistic electrons, so those electrons have not had as much time to diffuse along the local magnetic field lines. This jet-like source seen in Figures \ref{fig:SNTF1_C}, \ref{fig:SNTF1_Zoom_C}, and \ref{fig:SNTF1_X} could be the source of relativistic particles responsible for illuminating the filaments of SNTF1.

While SNTF1 is very similar to the morphology of the harps studied by \citet{Thomas2020}, we do not see a similar morphology in SNTF2 or SNTF3. In particular, there are two differences in the morphologies of SNTF2 and SNTF3 that do not align with the morphologies of the harps and SNTF1. First, the filaments comprising SNTF2 and SNTF3 all have comparable lengths. Second, there are no compact radio sources detected which could serve as candidates for the origin of relativistic electrons near SNTF2 or  SNTF3.

\subsubsection{Possible Sources of Relativistic Electrons} 
It is important to note that the NTFs throughout the GC generally exhibit brightness enhancements somewhere along their lengths \citep{LaRosa2000,Yusef-Zadeh2022}, including SNTF2 and SNTF3. Furthermore, the source of relativistic particles may not appear in the radio images studied in this work. Indeed, the source of relativistic electrons illuminating the harps studied by \citet{Thomas2020} was not directly detected in that work. Therefore, the regions of enhanced brightness within the NTFs could reveal the segments along the lengths of the filaments that are closer to the source of the relativistic particles.

Indeed, multiple possible examples of brightness enhancements coinciding with the source of electrons are present within the NTFBs studied in this work. There is a clear brightness enhancement within the filaments of SNTF1 coinciding with the possible jet-like structure. Furthermore, the filaments within SNTF2 reach maximum brightness along segments where the Radio Shell structures seem to approach the filaments. These brightness enhancements could therefore indicate that the jet and the Radio Shell structures are generating the relativistic electrons illuminating SNTF1 and SNTF2 respectively. The Radio Shell would therefore be an extended source of relativistic electrons. An extended structure could produce relativistic electrons if it is independently magnetized. If the magnetic field of this structure undergoes reconnection with the magnetic field of the nearby NTFB relativistic electrons could be produced \citep{Guo2020}.

\subsubsection{NTFs Without Filament-length Gradients}
If the source of relativistic electrons is extended, or if multiple sources are present, there may not be an obvious gradient in NTFB filament length. For example, the Snake NTF has multiple brightness enhancements (knots) along its length (e.g. \citealt{Gray1995}). The multiple brightness-enhanced regions could indicate multiple sources of relativistic electrons, or multiple places (e.g. the kinks) where electrons are preferentially accelerated.

There are multiple possibilities for what an extended source of relativistic electrons in the GC could be. Recent MeerKAT images of the GC at 1 GHz have revealed a large-scale synchrotron structure which is possibly an outflow generated by a previous period of activity of Sgr A$\rm^*$ or a previous starburst \citep{Heywood2019}. This radio structure is also observed to be co-spatial with the ``chimneys'' seen at infrared and X-ray wavelengths, leading to the hypothesis that this structure is a bi-directional outflow of hot plasma within the GC \citep{Ponti2019,Ponti2021}. Most of the NTFs are observed within a region having the same scale as the chimneys, with some of the prominent NTFs like the Radio Arc and the Snake present at the very edge of this space \citep{Heywood2019,Ponti2021}. The spatial coincidence of these structures and the NTFBs indicates they could be contributing to the formation process of the NTFBs. In addition, the non-thermal nature of the chimneys indicates that it is a magnetized structure, meaning that this structure could be producing relativistic electrons through magnetic field reconnection as discussed for the Radio Shell.

In light of the localized brightness enhancements seen in many members of the GC NTFB population, either the CR diffusion model of \citet{Thomas2020} or magnetic reconnection are possible explanations of how the NTFs throughout the GC are powered depending on whether the source of the relativistic electrons is compact or extended.

\section{CONCLUSIONS} \label{sec:conc}
We observed three NTFBs at southern latitudes using the VLA in both C- and X-band, providing a frequency range from 4 - 12 GHz. These high-resolution observations allowed us to study the total-intensity properties of these structures in detail. The properties of our NTFB targets were compared with those of the larger NTF population to determine whether they are distinct populations or not. The key results of this work are summarized here:

\begin{enumerate}

    \item Our total-intensity distributions reveal the sub-filamentation of the SNTFs, with each one consisting of $\rm\geq$10 individual filaments. The filaments within the STFs tend to clump together as seen for SNTF1 and SNTF2, a property which has been observed previously in the Radio Arc \citep{Pare2019}. The somewhat more isolated Quill and Wishbone NTFBs are observed to bifurcate into separate filaments along their lengths, although they are not comprised of as many individual filaments as are the SNTFs.
    
    \item Shallower spectral indices for our NTFBs are observed than what is reported for the larger NTF population studied by \citet{Yusef-Zadeh2022}. We generally do not observe spectral index gradients along the filament lengths of our NTF targets with a notable exception being the convincing spectral index gradient observed for the Wishbone. The Wishbone has a shallower spectral index at lower latitudes coinciding with where the individual filaments comprising the NTFB seem to converge.
    
    \item We obtain minimum magnetic field strength estimates for our NTFBs derived from equipartition calculations on the order of 100$\rm\mu$G. These minimum magnetic field estimates are in agreement with the magnetic field strength estimates derived for other NTFs in the GC. 
    
    \item SNTF1 is possibly well characterized by the CR transport model developed by \citet{Thomas2020}. The jet-like source near SNTF1 is possibly the origin of relativistic electrons. Alternatively, the jet-like source could possess its own magnetic field which undergoes reconnection with the SNTF1 magnetic field. The CR transport model developed in \citet{Thomas2020} could possibly explain how many of the NTFs throughout the GC are powered, especially when considering the existence of numerous filamentary structures recently detected by MeerKAT \citep{Heywood2019}. Other NTFs, like the Radio Arc, which do not exhibit filament length gradients could instead by powered by CR production caused by magnetic reconnection resulting from nearby extended structures (such as the Radio Shell observed near the Radio Arc or the non-thermal chimneys).
    
    \item We find the NTFBs observed in this work and the larger NTF population share a number of properties, with the only systematic difference being the number of individual filaments comprising these structures. 
\end{enumerate}
In this work we have focused on the total intensity properties of the NTFB targets. It is also important, however, to analyze the polarimetric properties of the NTFBs and compare them to the polarimetric properties of the larger NTF population. This analysis will allow us to determine whether the NTFBs have the same polarization properties of the larger NTF population, and whether the detailed polarization data can provide insight for how the relativistic electrons are produced. The analysis of the polarized intensity, RM, and intrinsic magnetic field distributions of the NTFBs discussed in this paper will be presented in a subsequent paper.


\software{
    MIRIAD, \citep{Sault1995},
    Astropy \citep{Greenfield2014},
    CASA \citep{McMullin2007},
    Matplotlib \citep{Hunter2007}
    }

\bibliographystyle{aasjournal}
\bibliography{astronomy}

\end{document}